# Cutting corners to suppress high-order modes in Mie resonator arrays

Zaid Haddadin[1], Shahrose Khan[2], and Lisa V. Poulikakos*[2,3]

[1]Department of Electrical & Computer Engineering, UC San Diego

[2]Department of Mechanical & Aerospace Engineering, UC San Diego

[3]Materials Science & Engineering Program, UC San Diego

**ABSTRACT:** Mie resonators as lattice resonant metasurfaces have the capability to produce structural colour. However, design criteria for these metasurfaces are still being investigated. In this work, we numerically examine how the two-dimensional nanostructure shape in a lattice array affects the colorimetric response of the metasurface under linearly polarised light excitation. First, the transformation from a square-shaped to rectangle-shaped nanostructure array resulted in polarisation-sensitive metasurfaces with colorimetric outputs bound along a line on the CIE 1931 2-degree Standard Observer colour space. The bounds of the colorimetry line were tuneable to any desired chromatic range. Second, the removal of the corners in square- or rectangle-shaped nanostructures to create t-shaped nanostructure arrays displayed a dampening effect on the high-order resonance. Finally, we analytically determined that the colour saturation could increase when moving from rectangle-shaped to t-shaped nanostructure arrays. From these results, we present two design guidelines for lattice resonant metasurfaces: (1) Constructing the nanostructure to support fundamental resonances at different wavelengths enables two-colour-bound movement when excited by successive angles of linearly polarised light; (2) Removing portions of the nanostructure that only support high-order resonances dampens these modes while maintaining support for fundamental resonances. These results present first-principles guidelines for engineering nanoparticles in lattice resonant metasurfaces, offering a new toolbox for polarised-light sensing and colorimetric applications.

TOC Graphic

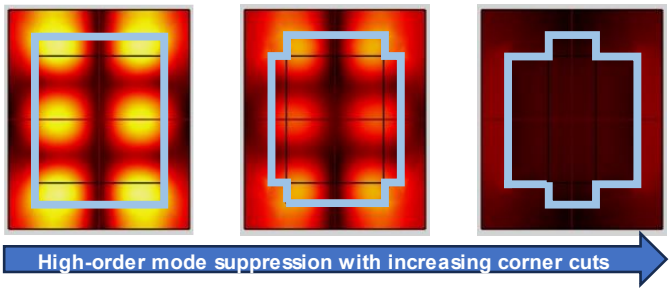

High-order mode suppression with increasing corner cuts

## Introduction

Structural colour arises from the scattering, diffraction, and dispersion of light by interactions with subwavelength structures[1–8], which result from interference effects between electromagnetic waves and matter[9,10]. Other mechanisms for colour generation include light absorption, transmission, and reflection by pigments[11–13], which can exhibit instability[14–17]; and light emission through luminescence[18–20], which can be limited in duration and intensity[20].

Structural colours offer vivid saturations with enhanced resolution, tunability, and durability compared to their chemical counterparts[10]. They have been studied since initial observations by Robert Hooke[21] and Isaac Newton[22]. Research has explored the generation of structural colours for various applications: biomedical sensing[23–32], coloured displays[32–34], solar energy conversion[35–37], information encryption[38,39], and more[40–42]. However, full realisation of these applications necessitates a deeper understanding of the mechanisms enabling control and tunability of colour saturation, resonant wavelength, and colorimetric multiplexing.

Metasurfaces – engineered periodic arrangements of sub-wavelength nanostructures – offer efficient, tuneable systems for generating structural colour. These colorimetric metasurfaces can consist of (i) plasmonic[43–46], (ii) dielectric[47–51], or (iii) hybrid[52–60] material systems. Plasmonic systems – including hybrid systems with plasmonic materials – suffer high optical losses in the visible light regime, limiting production of saturated structural colours[61–63]. In contrast, dielectric materials, which can exhibit low dissipative losses and medium-to-high refractive indices, serve as efficient generators of structural colour[64–66].

Dielectric structures can support Mie resonances, which depend on the size and geometry of the particles interacting with incident electromagnetic waves[67–70]. Structural colour can be achieved by tuning the geometry of the dielectric particle to confine the resonance within the visible light regime[71]. The effect was attributed to magnetic dipole excitation in 2012[67,68]. Since then, research has explored how arrays of dielectric nanostructure elements, known as all-dielectric metasurfaces, can achieve colour filtering[72,73].

All-dielectric colorimetric metasurfaces have been demonstrated using one-dimensional subwavelength features, such as gratings[74], or using two-dimensional subwavelength features in the form of periodically arranged nanoparticles[75]. In the latter case, the periodic repetition of subwavelength Mie scatterers enhances the magnetic dipole excitation compared to the single particle case, because of the collective resonant nature of a lattice array[76-86]. Alterations to the periodicity of the lattice array through structural adjustments of the particles and/or changes to the spacing among particles can modify the resonant wavelength[84-86].

This lattice resonant effect has been utilised to generate structural colours in reflectance using square-shaped nanostructure arrays made of silicon nitride[47]. Although a dampening effect on high-order Mie resonances was observed while preserving the fundamental resonance, the presence of high-order resonances in structures generating red colour was still reported[47]. Consequently, the design of nanostructure arrays that eliminate support for high-order Mie resonances remains an unanswered question.

This work introduces new criteria for the design of nanostructure arrays aimed at dampening or extinguishing high-order resonances. Finite element method simulations were conducted to study square- and rectangle-shaped nanostructure arrays. The rectangle-shaped nanostructure arrays demonstrated two distinct structural colour outputs in reflectance when excited by linearly polarised light aligned with either the vertical or horizontal axis; this is attributed to the separate Mie resonances arising from the different periodicities along those axes[83,86]. When the linearly polarised light was inclined towards or away from these orthogonal axes, a hybrid signal comprising amplitude-adjusted components of vertically and horizontally polarised light was observed. The resulting structural colour output of these hybrid signals was plotted along a line, bounded by two colours, on the CIE 1931 2-degree Standard Observer colour space[87,88]. The bounds of the colorimetry line were shown to be tuneable to any desired range, such as but not limited to blue-green, blue-red, and green-red.

The examination of near-field effects in the lattice arrays led to the hypothesis that removal of specific sections in the nanostructures that solely contribute to high-order resonances would dampen or extinguish them. In this work, the corners of the nanostructures met this criterion and removing them led to the predicted outcome. When the corners were removed without destabilising the fundamental resonances, the analytically calculated colour saturations increased. These findings emphasise how the intentional design of nanoparticles to selectively suppress or enable resonant modes offers an additional degree of freedom for tuning metasurfaces in polarised light-sensing applications.

## Results & Discussion

Lattice arrays consisting of silicon nitride particles on a silicon dioxide substrate were modelled in COMSOL Multiphysics® v5.6[89], a finite element method simulation software, using the Wave Optics module[90] and solved at the

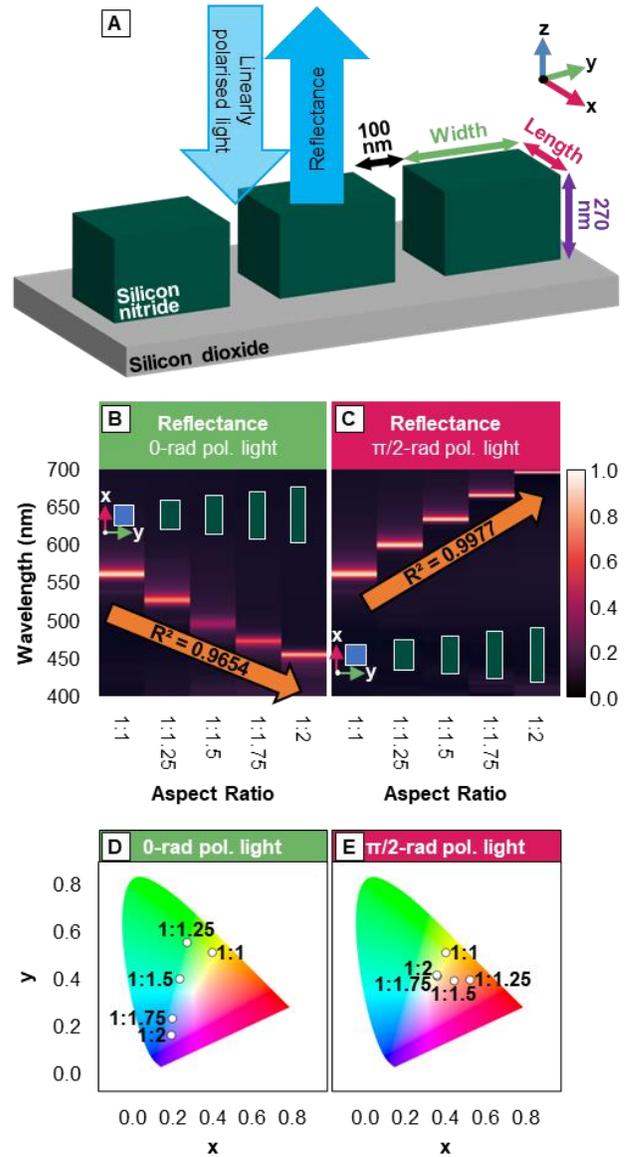

**Figure 1. (A)** Schematic of lattice resonant metasurface comprised of rectangle-shaped, silicon nitride nanostructures on a silicon dioxide substrate. **(B,C)** Heatmaps showing reflectance spectra of varying aspect ratios of rectangle-shaped nanostructure arrays acquired by excitation with *(B)* 0 rad (y-axis) or *(C)* π/2 rad (x-axis) linearly polarised light, illustrating the linear relationship between width or length of the nanostructure and the resonant wavelength. **(D,E)** Colorimetric outputs predicted from each reflectance spectra in (B) and (C) plotted on the CIE 1931 2-degree Standard Observer colour space, illustrating movement from green-to-blue when shortening the rectangle width and green-to-red when increasing the rectangle length.

San Diego Supercomputer Center[91]. These arrays were designed with Floquet periodicity conditions (see Supplementary Material, Section S1 for implementation details). Each particle had a fixed height of 270 nm, and the gap among particles was held at 100 nm (Figure 1A). The two-dimensional shape, width, and length of the nanostructure were varied across arrays. The tested shapes included rectangles and t-shaped configurations, where t-shapes represent rectangles with the corners removed. The periodici-ty along the y-axis (or x-axis) was calculated by [width (or length) + 100 nm]. Linearly polarised light was used to excite the nanostructure arrays at angles of 0 rad (y-axis polarised), π/6 rad, π/4 rad, π/3 rad, and π/2 rad (x-axis polarised). Reflectance data and near-field electric and magnetic field norm data were obtained for each study using the built-in functions of COMSOL Multiphysics® v5.6[89]. The generated reflectance spectra were utilised to predict the structural colour output on the CIE 1931 2-degree Standard Observer colour space[87,88] (see Supplementary Material, Section S2 for colour predicting calculations). These calculated structural colour outputs were converted to the CIE L*a*b* colour space[92,93] to determine the colour saturation as a ratio of chroma to lightness[94-97] (see Supplementary Material, Section S3 for colour saturation calculations).

An increase in aspect ratio from 1:1 to 1:2 with constant volume in rectangle-shaped nanostructure arrays (see Supplementary Material, Section S4 for parameters) resulted in a linear blue-shift in resonance location when excited with linearly polarised light at the 0 rad angle (Figure 1B) and a linear red-shift in resonance location when excited with light polarised at the π/2 rad angle (Figure 1C). These linear blue- and red-shifts had $R^2$ values of 0.9654 and 0.9977, respectively. These shifts corresponded to chromatic movements from green to blue in the former case (Figure 1D) and from green to red in the latter case (Figure 1E) on the CIE 1931 2-degree Standard Ob-server colour space[87,88].

When maintaining a constant aspect ratio of 1:1.25 and exciting the nanostructure with linearly polarised light oriented at the π/6 rad, π/4 rad, and π/3 rad angles, hybrid signals of the resonances from the 0 rad and π/2 rad polarisations were observed (Figure 2A). These hybrid signals were amplitude-adjusted dependent on the angular deviation from the 0 rad or π/2 rad axes. The colorimetric output of these spectra appeared on a colorimetric line bounded by the outputs of the 0 rad and π/2 rad polarisations (Figure 2B). Examinations of the near-field plots at the resonant wavelength revealed a Mie lattice resonance and a magnetic dipole resonance (Figures 2C and 2D): electric field norm plots showed a coupling between the nanostructure and the substrate (Figure 2C); magnetic field norm plots displayed coupling among the nanostructures and between the nanostructure and substrate (Figure 2D). This is in accord with results previously reported[47].

However, some volumes of rectangle-shaped nanostructures displayed additional, unwanted resonances in the lattice array configurations. For example, a rectangle-shaped nanostructure with dimensions of 327.4-nm-by-247.5-nm displayed two resonant peaks when excited by linearly polarised light at 0 rad (Figure 3A): a high-order resonance at 432 nm with amplitude 0.73 and a fundamental resonance at 572 nm with amplitude 0.49. Analysis of the electric field norms at the high-order and fundamental resonance wavelengths revealed that the corners of the nanostructure supported nodes belonging to the high-order resonance but not the fundamental resonance (Figure 3B). To destabilise the support of the high-order resonance, the corners of the rectangle-shaped nanostructures were cut to create our first t-shaped nanostructure array with arm thicknesses of 247.5 nm and 187.1 nm (Figure 3C, Inset). With this new structure, the high-order reso-nance in the reflectance spectrum was blue-shifted to 421 nm with a dampened amplitude of 0.38 (Figure 3C); the fundamental resonance blue-shifted to 567 nm with a maintained amplitude of 0.50. The near-field examinations at the high-order resonance wavelength showed a decrease in the maximum electric field norm value from 7.84E8 V/m with the rectangle-shaped nanostructure array to 6.58E8 V/m in the t-shaped nanostructure array (Figure 3D); the maximum electric field norm values at the fundamental resonance differed by 0.01E8 V/m. To completely suppress the high-order resonance, a larger corner cut had to be made to create a second t-shaped nanostructure array with arm thicknesses of 247.5 nm and 90.7 nm (Figure 3E, Inset). Herein, the high-order resonance disappeared (Figure 3E); the fundamental resonance blueshifted to 559 nm with an increased amplitude of 0.61. The maximum value of the electric field norm at the high-order resonance was 1.97E8 V/m (Figure 3F); the maximum value of the electric field norm at the fundamental resonance increased by 0.77E8 V/m, when compared with the rectangle-shaped nanostructure array result, to 4.39E8 V/m.

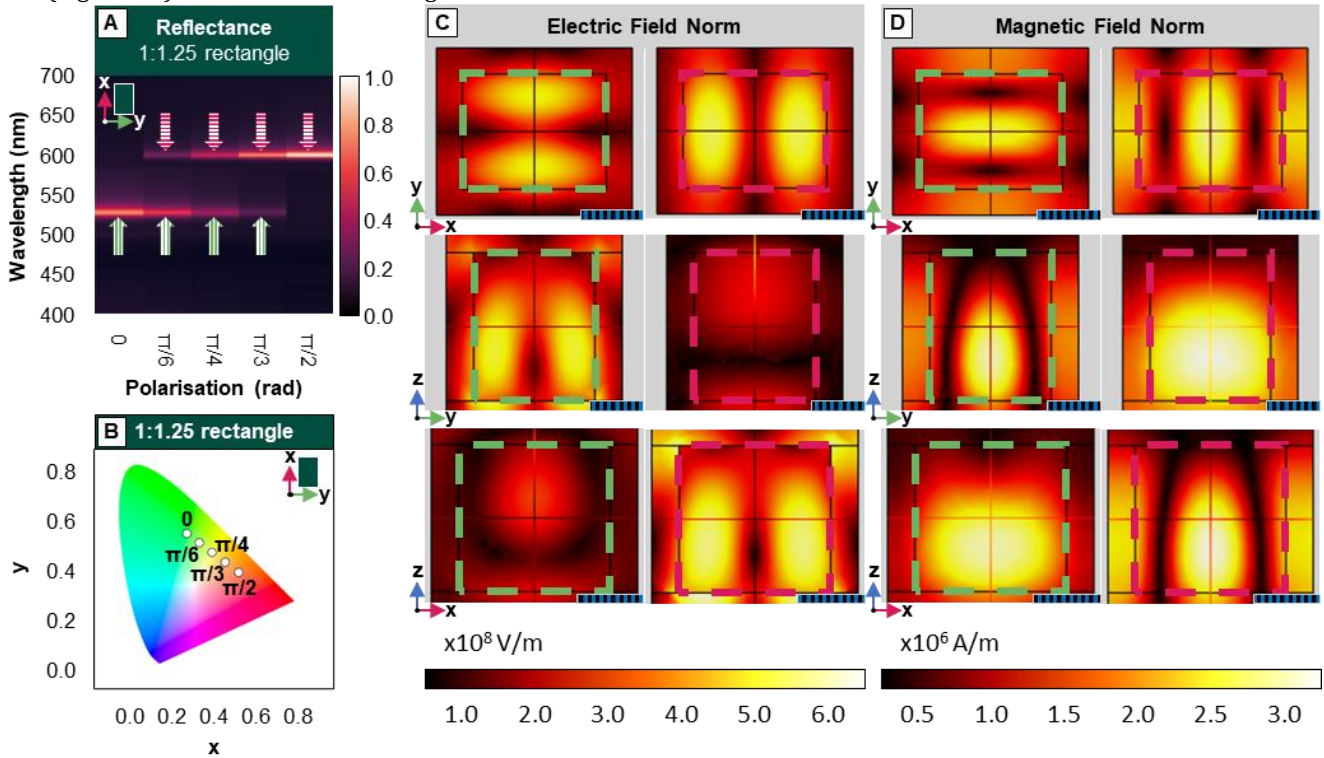

**Figure 2.** *(A)* Heatmap displaying reflectance spectra of a 1:1.25 aspect ratio rectangle excited by linearly polarised light at the 0 rad (y-axis), π/6 rad, π/4 rad, π/3 rad, and π/2 rad (x-axis) angles. Vertical, green-hashed arrows indicate the signal shared by the 0 rad polarisation resonance. Horizontal, red-hashed arrows indicate signal shared by π/2 rad polarisation resonance. *(B)* Colorimetric output predicted from each spectrum in (A) plotted on the CIE 1931 2-degree Standard Observer colour space. *(C,D)* Near-field *(C)* electric and *(D)* magnetic norm plots of the 1:1.25 aspect ratio rectangle excited by linearly polarised light along the *(Left Column)* 0 rad and *(Right Column)* π/2 rad angles. Dashed green or pink lines highlight the nanostructure boundary. Blue-and-black striped scale bars represent 100 nm.

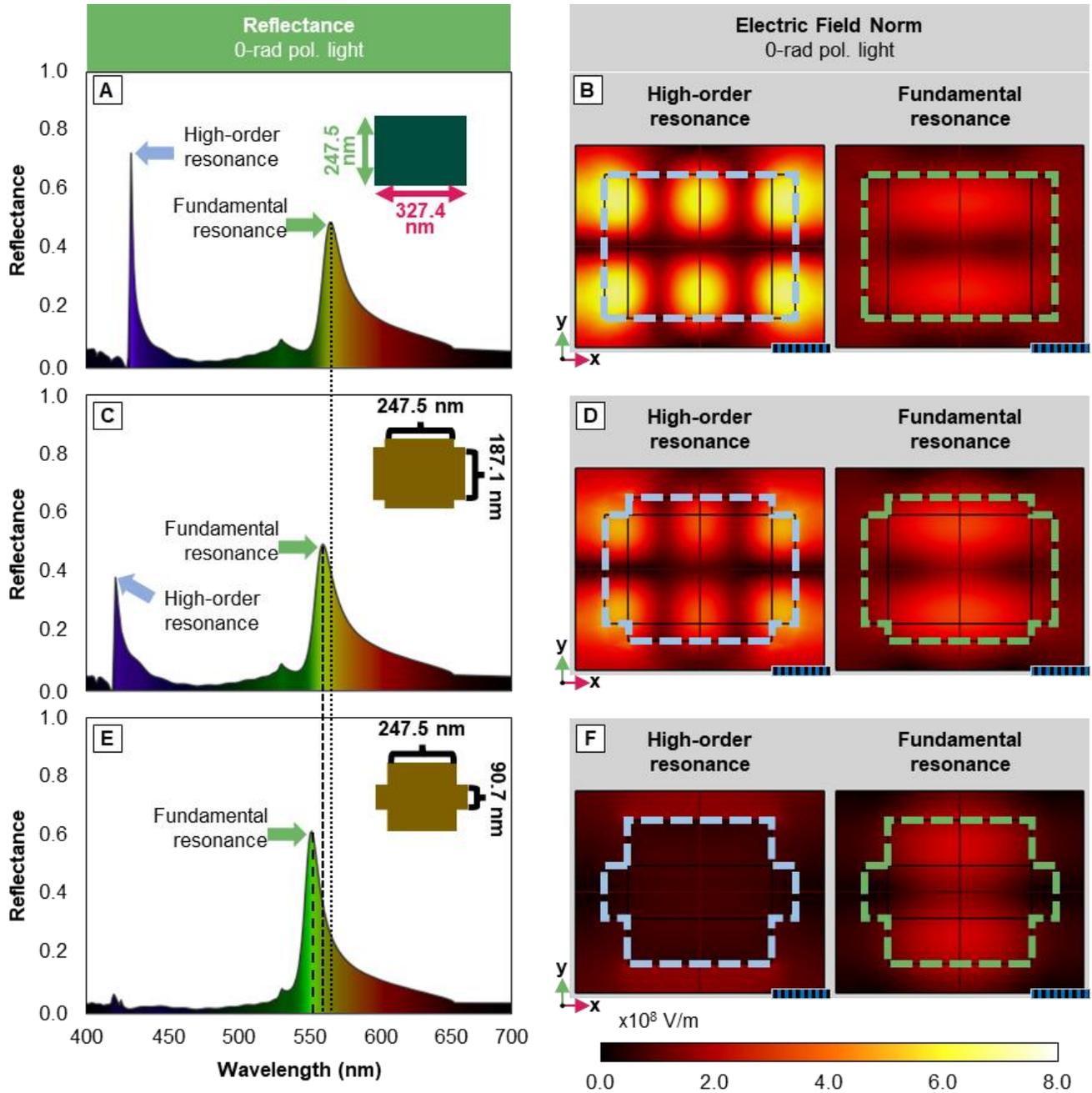

**Figure 3.** *(A,C,E)* Reflectance spectra of rectangle-shaped and t-shaped nanostructure arrays, indicated by insets, under linearly polarised light excitation at 0 rad (y-axis). Green arrows highlight the location of the fundamental resonance wavelength. Blue arrows highlight the location of the high-order resonance wavelength. The dotted, short dashed, and long dashed lines indicate the positions of the fundamental resonance wavelengths, which underwent a blue-shift when transforming a rectangle-shaped nanostructure array to a t-shaped nanostructure array. *(B,D,F)* Near-field plots of the electric norms of the *(Left Column)* high-order and *(Right Column)* fundamental resonances of the rectangle-shaped and t-shaped nanostructure arrays. Blue-and-black striped scale bars are 100 nm.

Overall, these cuts to the corners resulted in a suppressed high-order resonance, a blue-shifted fundamental resonance wavelength, an increased amplitude of the fundamental resonance wavelength, and an increased electric field norm magnitude of the fundamental resonance. We hypothesise the blue-shift observed for the fundamental resonance wavelength can be minimised with appropriate increases to the effective dimension along the axis of polarisation – in this case, periodicity along the y-axis. We also noticed decreases to the full width at half-maximum of the fundamental resonance wavelength as the corners were cut (see Supplementary Material, Sections S5 and S6 for reported full widths at half-maximum and more examples of this phenomenon). When combining this observation with the increase in amplitude of the fundamental resonance wavelength in the second t-shaped nanostructure array, we can hypothesise there exists a trade-off between the available bandwidths for resonance formation and the peak intensity due to the laws of conservation. In addition, this sharpened resonance due to the decreased full width at half-maximum can improve the saturation of the colorimetric output resultant from the reflectance spectrum.

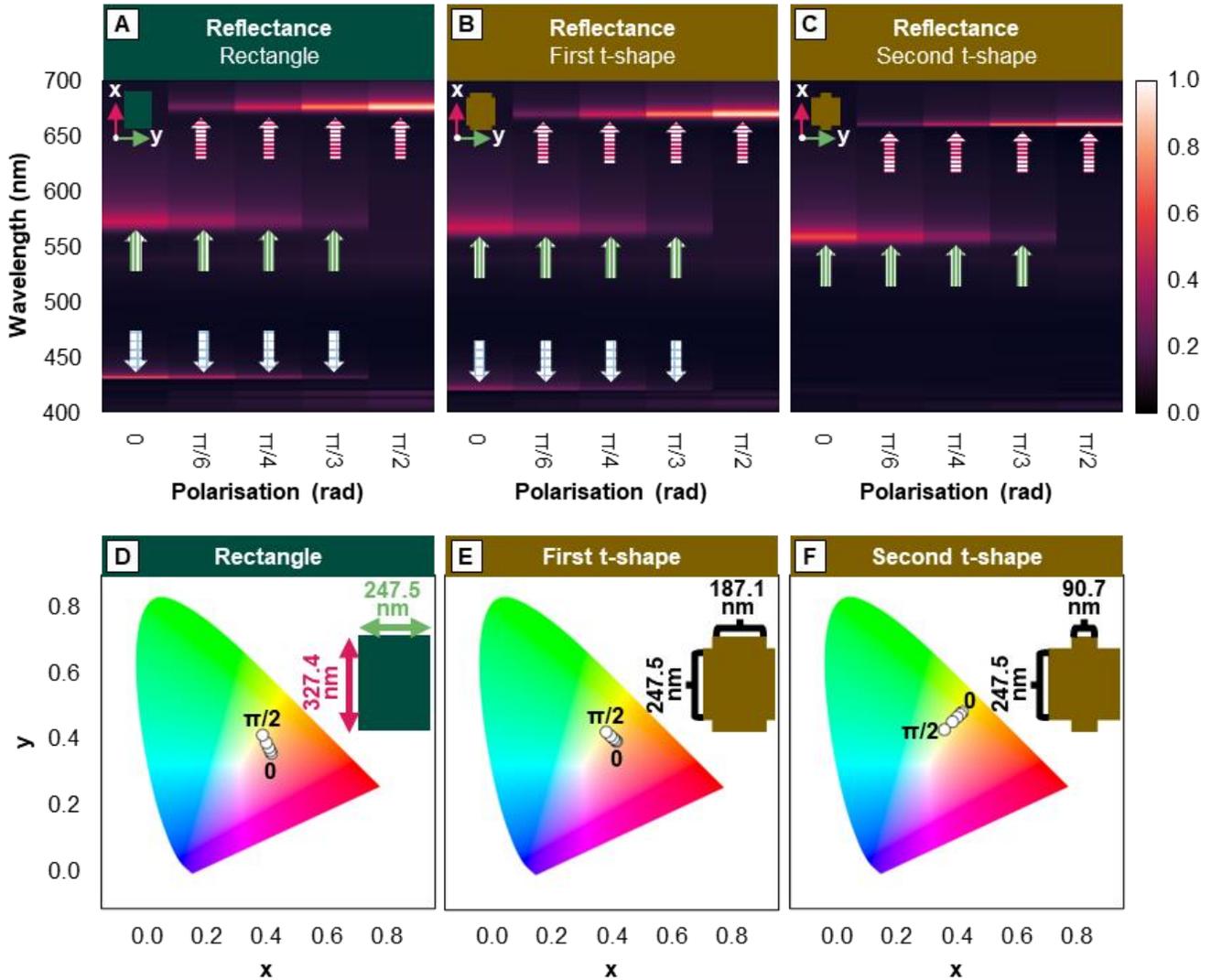

**Figure 4. (A-C)** Reflectance heatmaps acquired from the rectangle-shaped and t-shaped nanostructure arrays from Figure 3. Insets indicate the type of nanostructure. Blue, grid-filled arrows indicate the high-order resonances. Green, vertically striped arrows indicate fundamental resonance signals shared with the 0 rad (y-axis) polarisation. Red, horizontally striped arrows indicate fundamental resonance signals shared with the π/2 rad (x-axis) polarisation. The high-order resonances vanished with the t-shaped nanostructure array in (C). **(D-F)** Colorimetric output predicted from each reflectance spectrum in (A-C) plotted on the CIE 1931 2-degree Standard Observer colour space. Only the 0 rad (y-axis) and π/2 rad (x-axis) polarisations are labelled. All outputs are bounded along a line. The lines in (D, E) have a diagonal orientation. The line in (F) has an anti-diagonal orientation.

When exciting the rectangle-shaped, first t-shaped, and second t-shaped nanostructure arrays with linearly polarised light at the $\pi/6$ rad, $\pi/4$ rad, and $\pi/3$ rad angles, hybrid signals comprising the resonances from the 0 rad and $\pi/2$ rad polarisations were observed (Figure 4A-C). These hybrid signals were amplitude-adjusted based on angular deviation from the 0 rad or $\pi/2$ rad axes. The colorimetric output of these spectra on the CIE 1931 2-degree Standard Observer colour space[87,88] appeared as a line bounded between the outputs of the 0 rad and $\pi/2$ rad polarisations (Figures 4D-F). If the colorimetry lines in Figures 4D and 4E were classified as oriented diagonally, then the colorimetry line in Figure 4F can be described as oriented antidiagonally, closer to the edges of the colour space. Within this colour space, a tendency towards the edges represents a higher colour saturation[9,72], an indication that suppressing the high-order resonance impacts colorimetry. The colour saturation was calculated for all three lattice arrays when excited by linearly polarised light at 0 rad and $\pi/2$ rad (Figure 5). A greater range of increase in colour saturation was calculated under 0 rad polarisation compared to under $\pi/2$ rad polarisation when moving from the rectangle-shaped to the second t-shaped nanostructure arrays, an implication that suppression of the high-order resonance poses greater weighting on the final colour saturation than the decreased full widths at half-maximums of the resonance peaks. We chose to not conduct null hypothesis significant testing on the colour saturation calculations, because of the small sample sizes per category (n=4) and the numerical nature of this work provides a-priori knowledge that a colorimetric difference must exist if the resonances in reflectance across spectra are different from one another. Consequently, our evaluations and conclusions of the colour saturations results are qualitative.

## Conclusions

Based on the numerical studies conducted in this work, the following design criterion for nanostructures in lattice arrays can be identified: Removing sections of the nanostructure that solely contributed to high-order resonances dampened or extinguished those resonances while maintaining support for the fundamental resonances. This was established by investigating the near-field effects of the high-order and fundamental resonances of the rectangle-shaped nanostructure arrays. The high-order resonances exhibited activity at the corners and centres of the nanostructures, whereas the fundamental resonances only displayed activity at the centres of the nanostructures. Removal of the corners resulted in the dampening and extinction of the high-order resonance, as demonstrated in Figure 3.

This work reveals a new avenue for the design of nanostructures in lattice arrays. It holds potential in the development of two-colour-bound colorimetric sensors for incident linearly polarised light with rectangle-shaped-turned-t-shaped nanostructure arrays. And it can be extrapolated to the design of single-colour pixels using square-shaped-turned-t-shaped nanostructure arrays (see Supplementary Material, Section S6 for results and discussion). Moreover, t-shapes are not likely the sole type of shape transformation enabling the dampening of high-order resonances. Future research can explore optimisation techniques for selectively removing sections of nanostructures to destabilise high-order resonances while maintaining support for fundamental resonances. Our work removed all four corners of the rectangle-shaped structures, but investigating the effects of removing one, two, or three of the corners could yield new findings, especially when examining effects of asymmetric corner removal for creating chiral structures for the colorimetric sensing of incident clockwise and counterclockwise circularly polarised light without the need for circular dichroism setups[98]. Finally, a question that emerged during this work is whether it would be possible to alter the nanostructures to exclusively support a fundamental resonance along a single axis. Achieving this would enable the design of polarisation-sensitive colour filters using lattice arrays.

Altogether, it was possible to determine a new methodology for designing lattice arrays by examining near-field effects and removing portions of the nanostructure that support unwanted resonances.

## ASSOCIATED CONTENT

Supporting Information Available: The supplementary material text includes methods, calculations, additional data, details on the programming environments and scripts, and descriptions of all attached files. These files include the codes used to analyse the data of this work (see Supplementary Material, Section S7) and the data analysed in this published work (see Supplementary Material, Section S8). This material

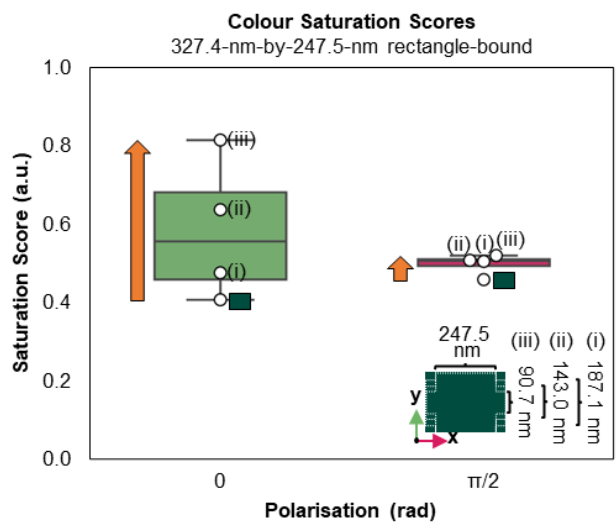

**Figure 5**. Box-and-whisker plot illustrating colour saturation measurements of rectangle-bound nanostructure arrays at 0 rad (y-axis) and $\pi/2$ rad (x-axis) linearly polarised light excitation. The green rectangle represents the control structure. (i), (ii), and (iii) indicate t-shape structures where (i) has the smallest area of corners removed and (iii) has the largest area of corners removed (exact dimensions provided in inset). Orange arrows indicate movement from the rectangle-shaped control to the t-shape with the greatest area of corners cut out. Boxes represent the quartiles, with the horizontal grey line indicating the median. Whiskers represent the range of the distribution.

is available free of charge via the Internet at 

## AUTHOR INFORMATION

### Corresponding Author

* L.V.P., University of California San Diego, La Jolla, CA, USA; lpoulikakos@ucsd.edu.

### Author Contributions

Following CRediT definitions: conceptualisation: Z.H.; data curation: Z.H., S.K.; formal analysis: Z.H., S.K.; funding acquisition: L.V.P.; investigation: Z.H., S.K.; methodology: Z.H.; project administration: Z.H.; resources: L.V.P.; software: Z.H., S.K.; supervision: L.V.P.; validation: Z.H., S.K.; visualisation: Z.H., L.V.P.; writing – original draft: Z.H.; writing – review & editing: Z.H., S.K., L.V.P.

All authors have given approval to the final version of the manuscript.

### Funding Sources

All authors gratefully acknowledge funding from the Arnold and Mabel Beckman Foundation (Beckman Young Investigator Award, Project Number: 30155266).

## ACKNOWLEDGMENT

The authors thank Ping Chu, Subha Sivagnanam, and Dong-Chu "DJ" Choi for helpful discussions and technical support.

**Cutting corners to suppress high-order modes in Mie resonator arrays**

*Supplementary Material*

Authors: Zaid Haddadin[1], Shahrose Khan[2], Lisa V. Poulikakos[*,2,3]

[1]Department of Electrical & Computer Engineering, UC San Diego

[2]Department of Mechanical & Aerospace Engineering, UC San Diego

[3]Materials Science & Engineering Program, UC San Diego

**Keywords:** lattice, resonances, metasurfaces, saturation, silicon nitride, structural colour

## Table of Contents

## Section S1: COMSOL Multiphysics® v5.6 Set Up

The COMSOL Multiphysics® v5.6 base software [1] and the Wave Optics Module [2] were used to for this study. The COMSOL Multiphysics® file was created with an "Electromagnetic Waves, Frequency Domain" layer in the three-dimensional (3D) space. The file is created with an empty study.

### S1.1: Global Definitions

Three sub-layers were created underneath the default "Global Definitions" layer: (i) "Nanostructure Parameters", (ii) "Model Parameters", and (iii) "Optics Parameters". The attributes and entries contained in these layers can be seen in Tables S1-S3

**Table S1.** "Nanostructure Parameters" attributes and entries. "Dep. Var." entries in the "Expression" and "Value" columns denote dependent variables that changed among simulations. By altering the *_depth and *_width attribute entries, it's possible to create square-, rectangle-, and t-shaped nanoparticles. (* = wildcard: any character or sequence of characters)

| Name | Expression | Value | Description |
|---|---|---|---|
| A_interresonator_gap | 100 [nm] | 1E-7 m | The distance between nanostructure A and an adjacent one. |
| A_nanostructure_depth | *Dep. Var.* | *Dep. Var.* | The depth of nanostructure A. |
| A_nanostructure_width | *Dep. Var.* | *Dep. Var.* | The width of nanostructure A. |
| B_interresonator_gap | 100 [nm] | 1E-7 m | The distance between nanostructure B and an adjacent one. |
| B_nanostructure_depth | *Dep. Var.* | *Dep. Var.* | The depth of nanostructure B. |
| B_nanostructure_width | *Dep. Var.* | *Dep. Var.* | The width of nanostructure B. |
| nanostructure_thickness | 270 [nm] | 2.7E-7 m | The thickness (or height) of the nanostructure. |

**Table S2.** "Model Parameters" attributes and entries. "Indep. Var." entries in the "Value" columns denote independent variables that changed among simulations.

| Name | Expression | Value | Description |
|---|---|---|---|
| domain_depth | lattice_depth_period | 4.274E-7 m | The depth of the entire domain. |
| domain_height | incident_height + transmitted_height + nanostructure_thickness | 4.27E-6 m | The height of the entire domain. This is composed of all parts of the model. |
| domain_width | lattice_width_period | 4.274E-7 m | The width of the entire domain. |
| incident_height | 2000 [nm] | 2E-6 m | The height of the domain through which light is incident towards the thin film. |
| lattice_depth_period | B_nanostructure_depth + B_interresonator_gap | *Indep. Var.* | The interval of distance between successive repetitions of the nanoresonator along the depth of the nanostructure. |
| lattice_width_period | A_nanostructure_width + A_interresonator_gap | *Indep. Var.* | The interval of distance between successive repetitions of the nanoresonator along the width of the nanostructure. |
| transmitted_height | 2000 [nm] | 2E-6 m | The height of the domain through which light is transmitted after passing through the thin film. |

**Table S3.** "Optics Parameters" attributes and entries. "Dep. Var." entries in the "Expression" and "Value" columns denote dependent variables that changed among simulations. By altering the angle_polarisation_* attribute entries, it's possible to become alter the angle of incident polarisation. (* = wildcard: any character or sequence of characters)

| Name | Expression | Value | Description |
|---|---|---|---|
| wavelength_step | 1 [nm] | 1E-9 m | The steps taken in incrementing/decrementing the wavelength when running the simulation. |
| wavelength_min | 400 [nm] | 4E-7 m | The minimum value the wavelength of light can take for this study. |
| wavelength_max | 700 [nm] | 7E-7 m | The maximum value the wavelength of light can take for this study. |
| angle_polarisation_y | *Dep. Var.* | *Dep. Var.* | The angle of the incident y-component of the polarised light. |
| angle_polarisation_x | *Dep. Var.* | *Dep. Var.* | The angle of the incident x-component of the polarised light. |

## S1.2: Component 1

### S1.2.1: Definitions

In the "Definitions" layer of "Component 1", a sub-layer is created: "Electric mode field amplitudes". Table S4 contains the attributes and entries of this sub-layer.

**Table S4.** "Electric mode field amplitudes" attributes and entries.

| Name | Expression | Unit | Description |
|---|---|---|---|
| E0x | –sin(angle_polarisation_x) | | The x-component of the electric field amplitude. |
| E0y | cos(angle_polarisation_y) | | The y-component of the electric field amplitude. |

### S1.2.2: Geometry 1

In the "Geometry 1" layer of "Component 1" four sub-layers are added: (i) "Block" layer renamed to "Domain"; (ii) "Work Plane" layer renamed to "Nanostructure Work Plane"; (iii) "Extrude" layer; and (iv) "Form Union" layer.

(i) The settings for "Domain" layer can be seen in Figures S1.

(ii) The settings for "Nanostructure Work Plane" layer can be seen in Figure S2. Moreover, the "Nanostructure Work Plane" by default is made with a "Plane Geometry" sub-layer. Under the "Plane Geometry" sub-layer, three more layers are made: (a) "Rectangle" layer renamed to "Rectangle A" (settings in Figure S3, Left); (b) "Rectangle" layer renamed to "Rectangle B" (settings in Figure S3, Right); and (c) "Union" layer renamed to "Cross Nanostructure" (settings in Figure S4).

(iii) The settings for "Extrude" layer can be seen in Figure S5.

(iv) The settings for "Form Union" layer were set to "Form a union" for "Action:" and "Automatic" for "Repair tolerance:".

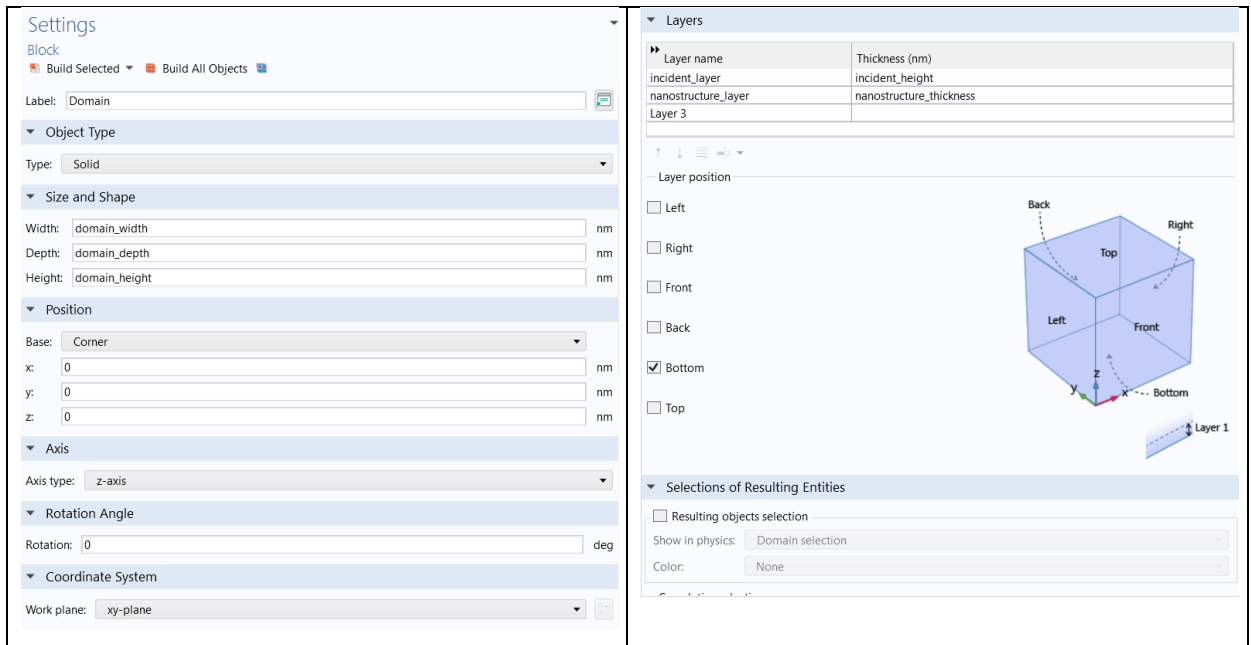

**Figure S1.** The settings used for the "Domain" sublayer of "Geometry 1".

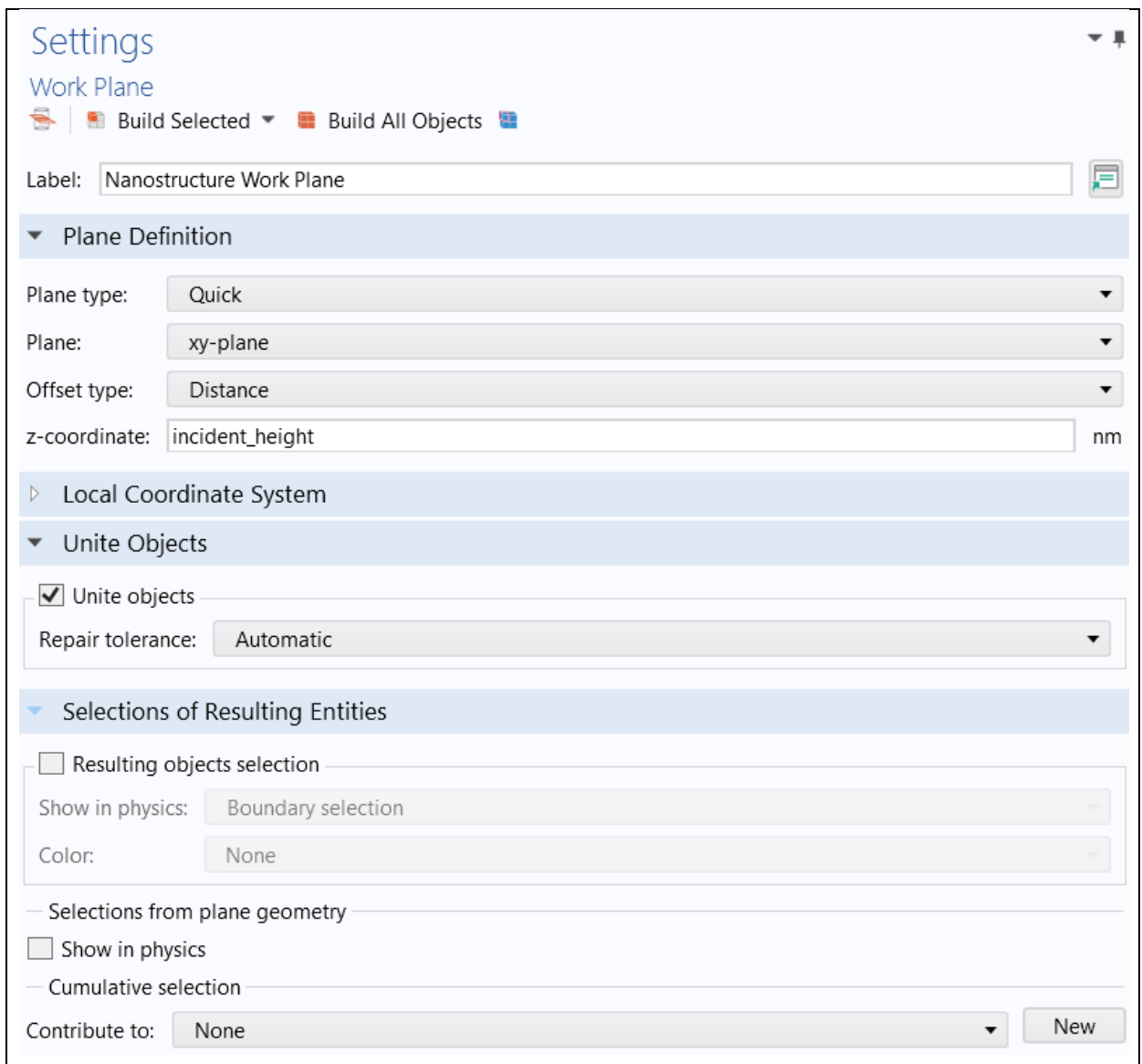

**Figure S2.** The settings used for the "Nanostructure Work Plane" sublayer of "Geometry 1".

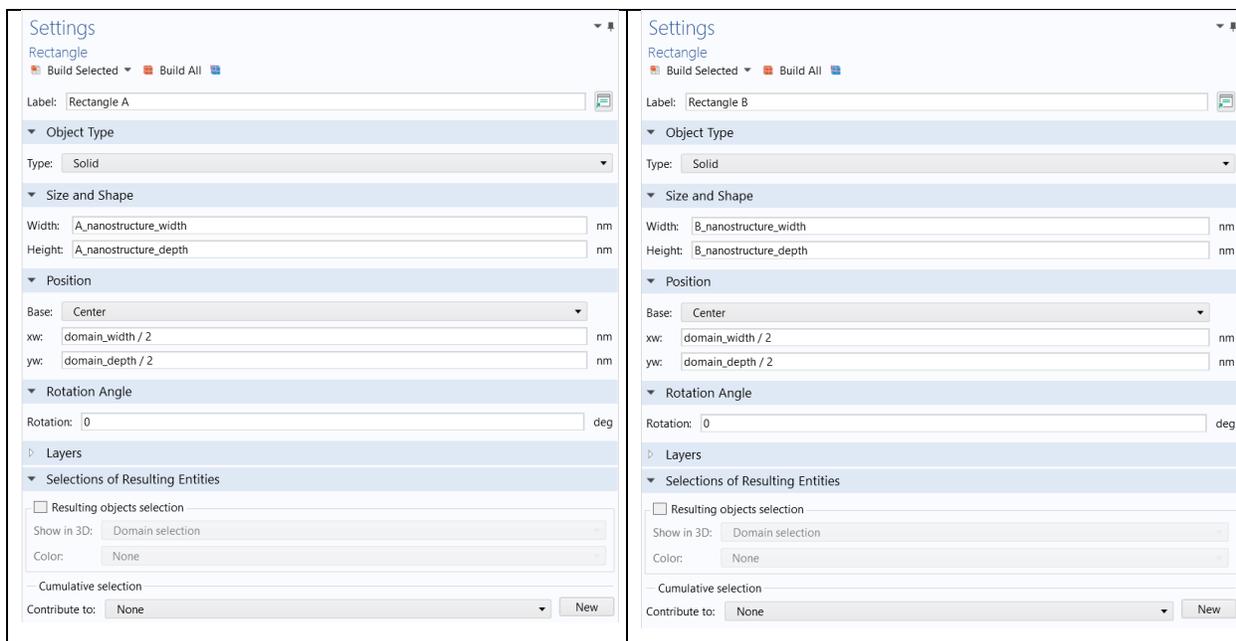

**Figure S3.** The settings used for **(Left)** "Rectangle A" and **(Right)** "Rectangle B" sub-layers of "Nanostructure Work Plane".

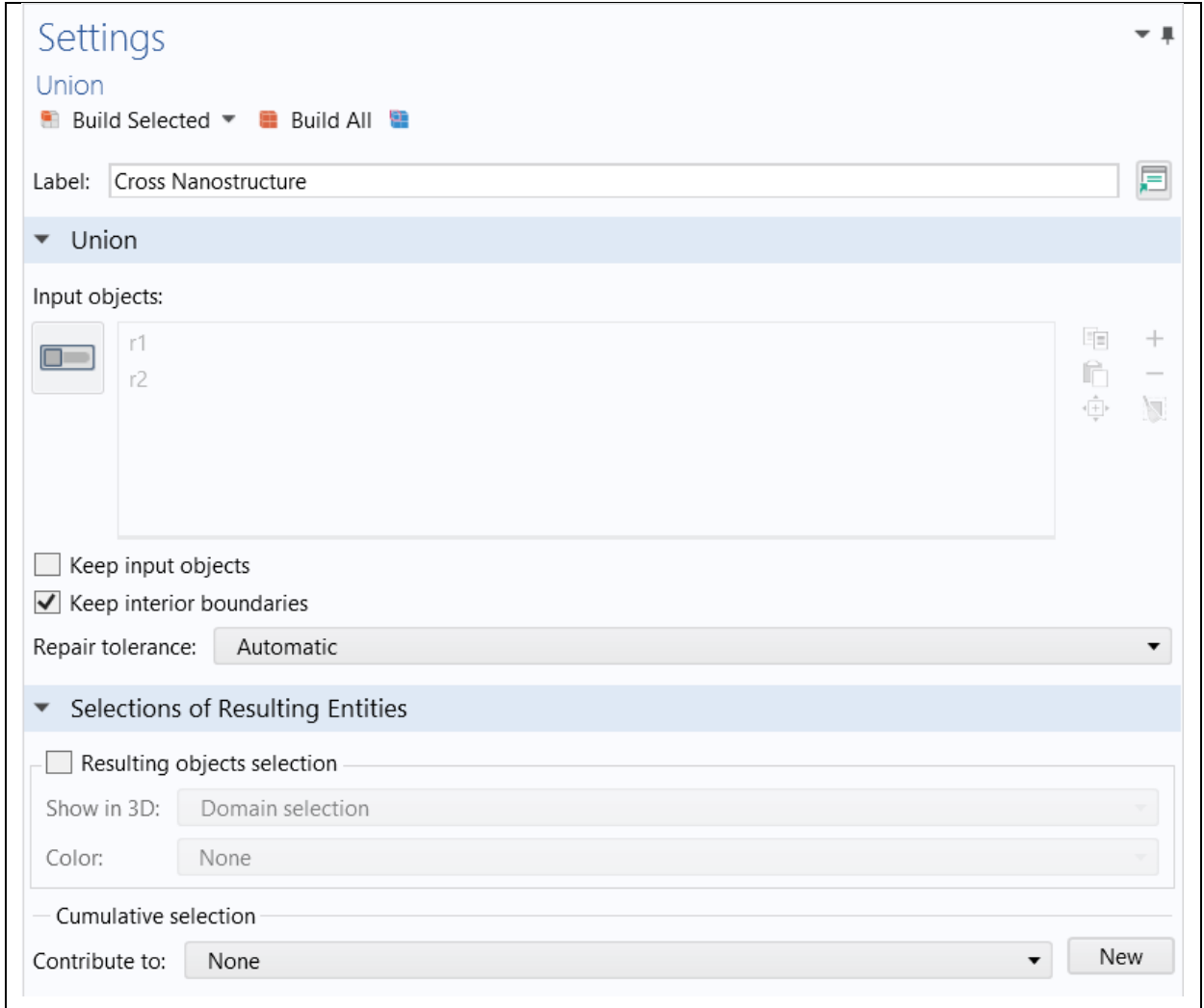

**Figure S4.** The settings used for the "Cross Nanostructure" sub-layer of "Nanostructure Work Plane". The "r1" and "r2" objects in the settings refer to the "Rectangle A" and "Rectangle B" sub-layers, respectively.

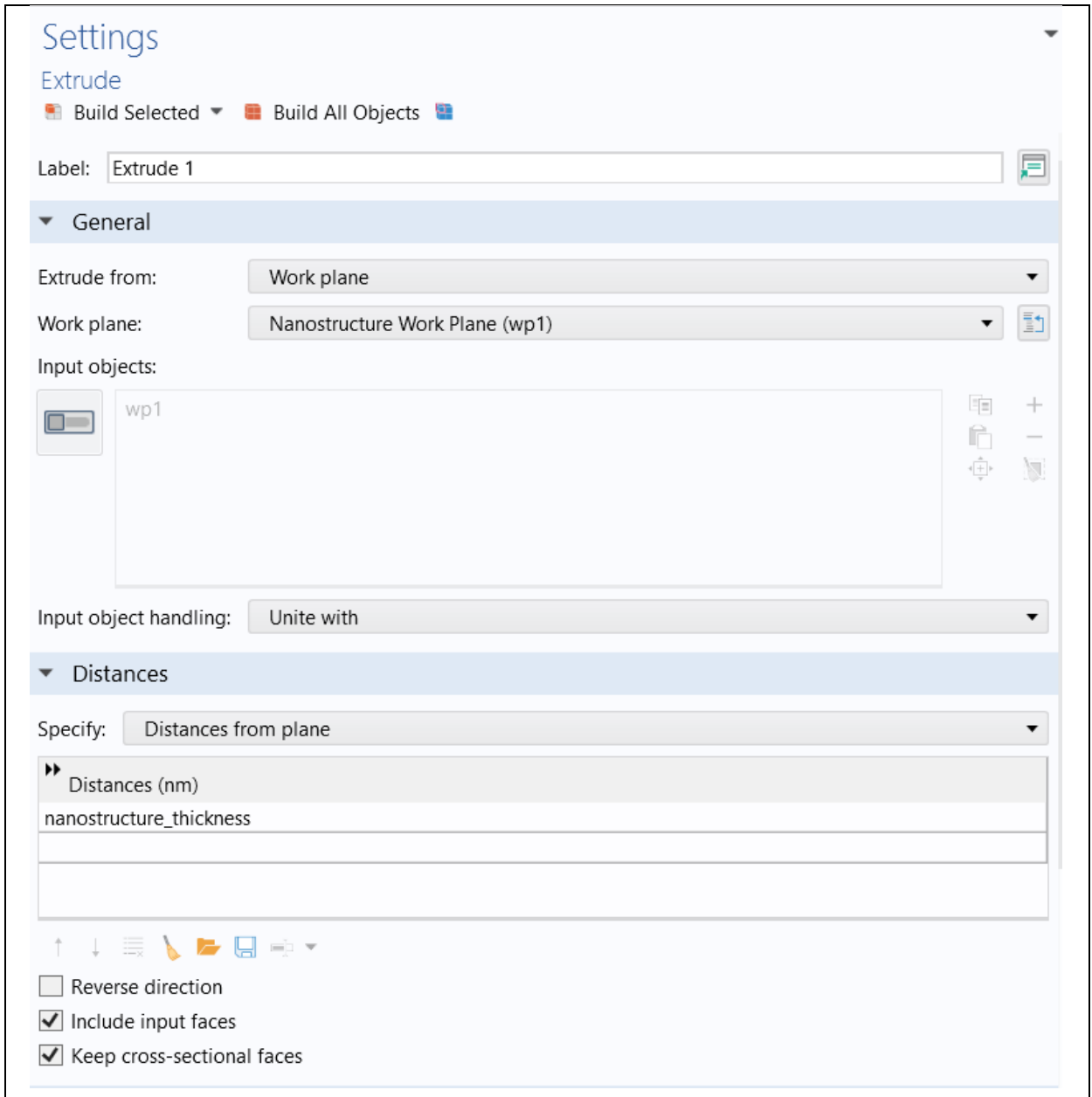

**Figure S5.** The settings used for the "Extrude" layer of "Geometry 1". The "wp1" object in the settings refer to the "Nanostructure Work Plane" layer.

### S1.2.3: Materials

Three materials were added using the in-built COMSOL Multiphysics® v5.6 Wave Optics Module library: (i) "Air (Ciddor 1996: n 0.23-1.690 μm)" [3]; (ii) "Si3N4 (Silicon nitride) (Luke et al. 2015: n 0.310-5.504 μm)" [4]; and (iii) "SiO2 (Silicon dioxide, Silica, Quartz) (Ghosh 1999: α-Quartz, n(o) 0.198-2.0531 μm)" [5]. Figure S6 shows the distribution of the materials in the model.

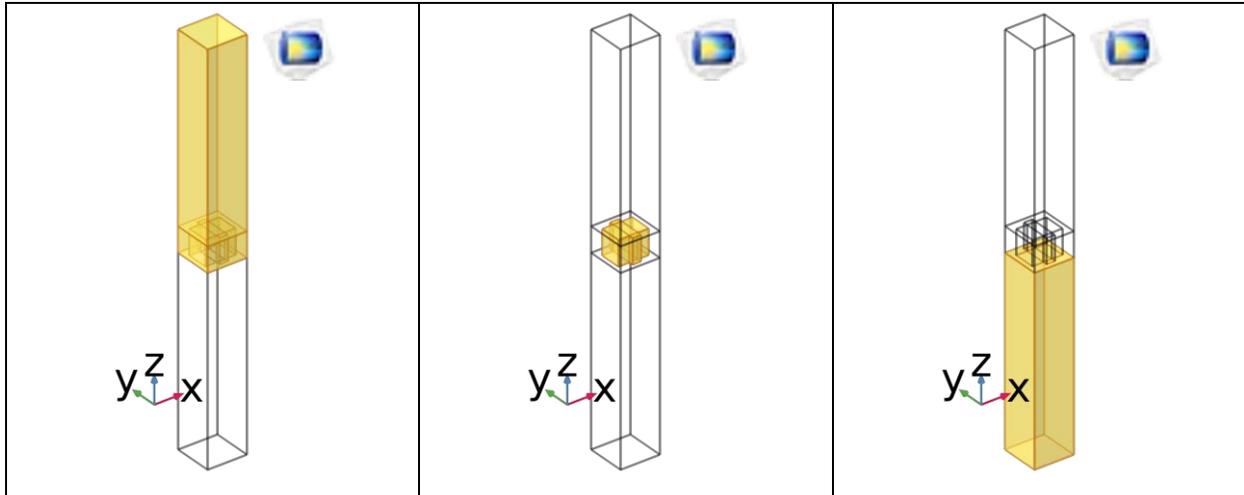

**Figure S6.** The yellow highlights represent the distribution of *(Left)* air, *(Middle)* silicon nitride, and *(Right)* silicon dioxide in the model. Air encompasses domains 2-3. Silicon nitride encompasses domains 4-8. Silicon dioxide encompasses domain 1. Images provided using the COMSOL Multiphysics® in-built functionality.

### S1.2.4: Electromagnetic Waves, Frequency Domain

The "Formulation" of the "Electromagnetic Waves, Frequency Domain" layer is set to "Full field". By default, three sub-layers are provided with this layer: (i) "Wave Equation, Electric 1"; (ii) "Perfect Electric Conductor 1"; and (iii) "Initial Values 1". Figure S7 shows the settings of "Wave Equation, Electric 1" and "Initial Values". "Perfect Electric Conductor 1" is overridden by future settings; thus, the settings are not provided in this write-up.

Moreover, four additional sub-layers are created: (iv) "Port" layer named "Port 1"; (v) "Port" layer named "Port 2"; (vi) "Periodic Condition" layer named "Periodic Condition 1"; (vii) "Periodic Condition" layer named "Periodic Condition 2".

(iv) "Port 1" layer settings are provided in Figure S8. The "Add Diffraction Orders" at the bottom of the settings was selected. This added a "Orthogonal Polarization 1" layer under "Port 1" and under "Port 2", and sixteen "Diffraction Order" layers under "Port 2".

(v) "Port 2" layer settings are provided in Figure S9.

(vi) "Periodic Condition 1" layer settings are provided in Figure S10, Left.

(vii) "Periodic Condition 2" layer settings are provided in Figure S10, Right.

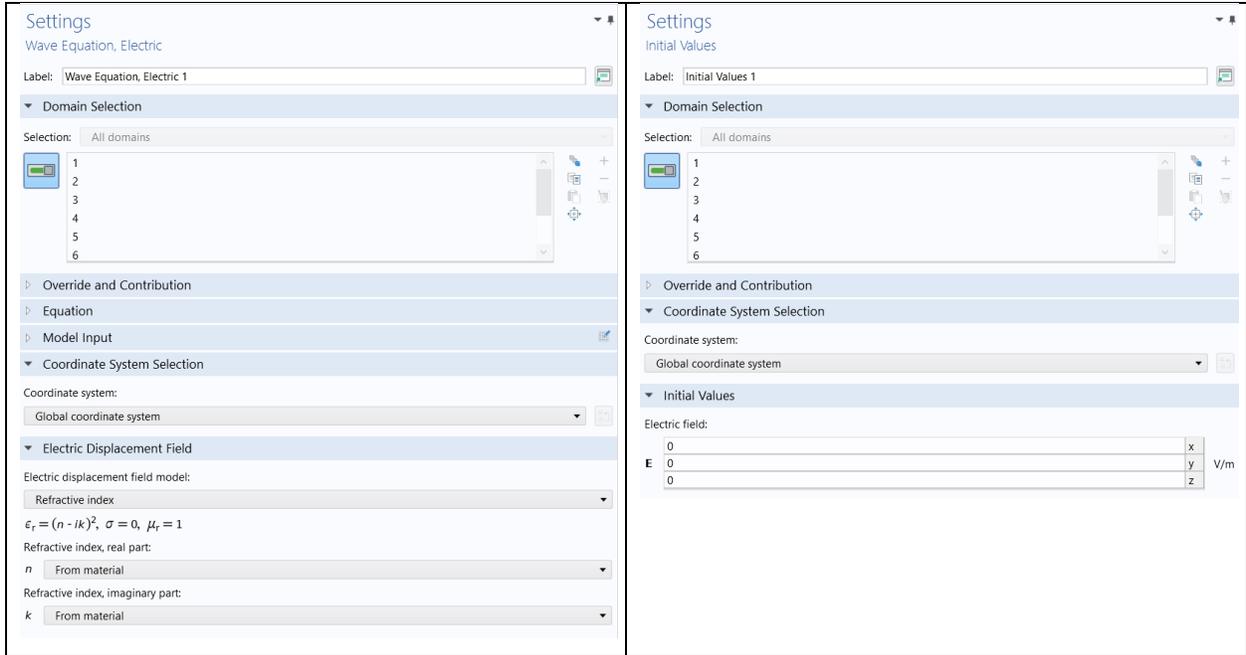

**Figure S7.** Settings for the *(Left)* "Wave Equation, Electric 1" layer and *(Right)* "Initial Values 1" layer of the parent "Electromagnetic Waves, Frequency Domain" layer.

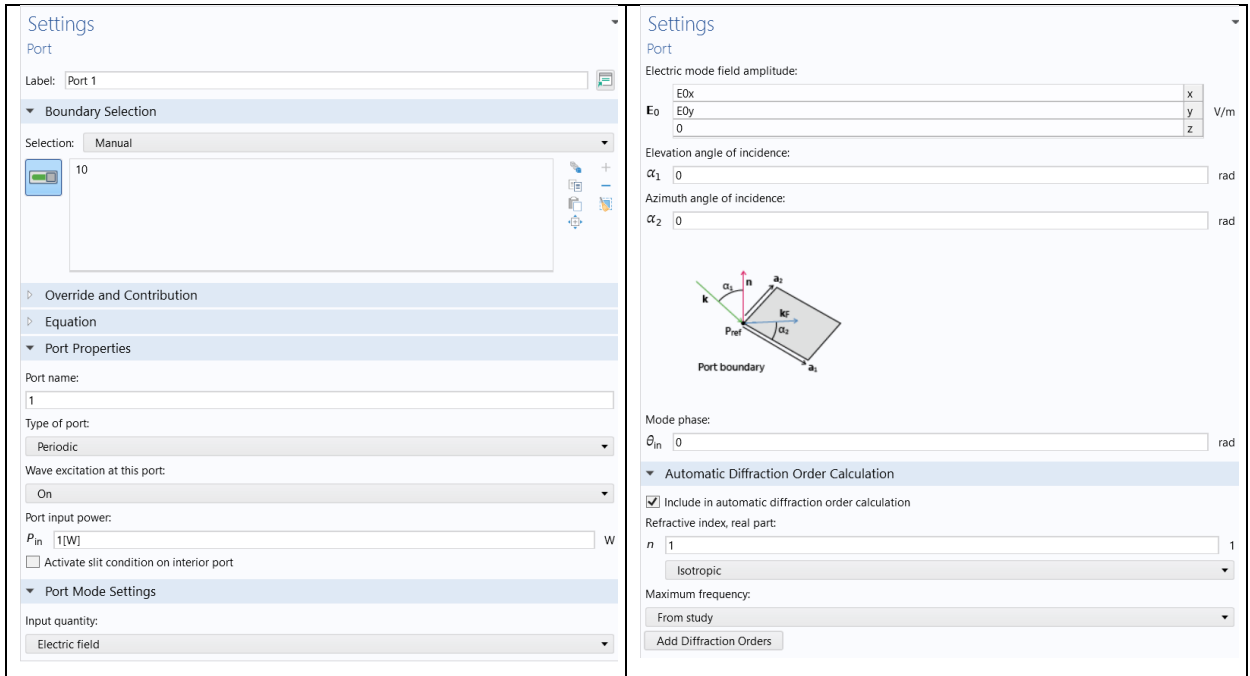

**Figure S8.** Settings for the "Port 1" layer of the "Electromagnetic Waves, Frequency Domain" layer.

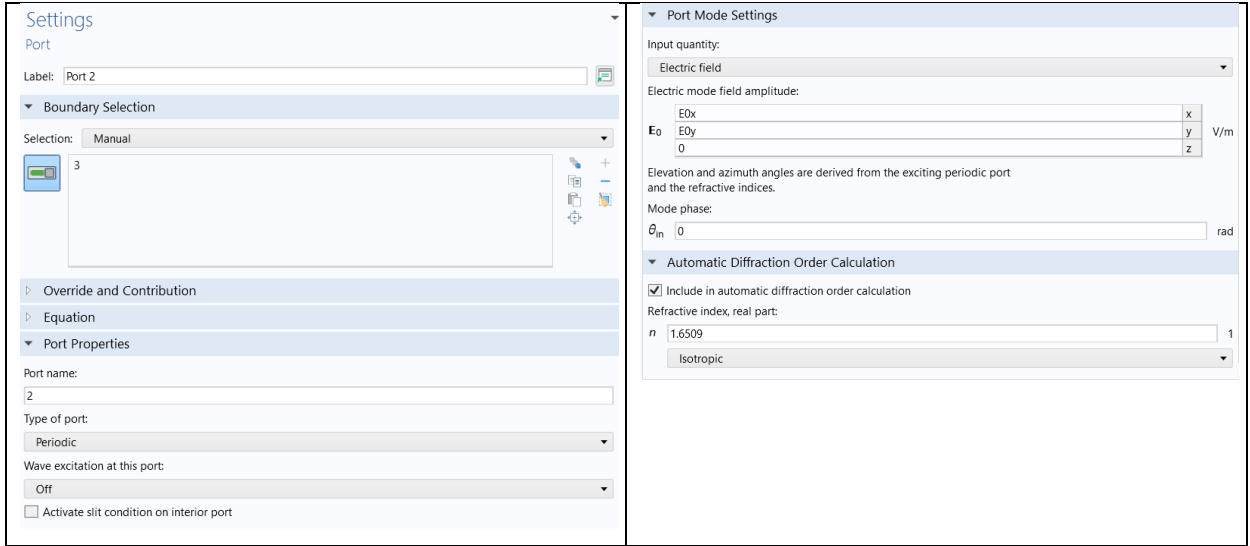

**Figure S9.** Settings of "Port 2" layer of the "Electromagnetic Waves, Frequency Domain" layer.

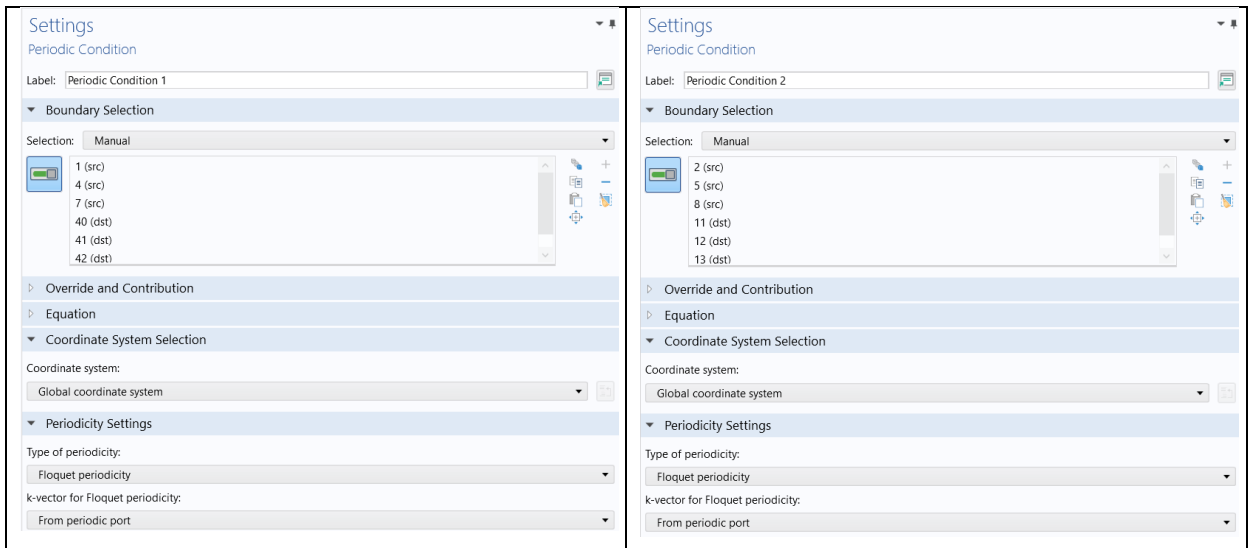

**Figure S10.** Settings for the **(Left)** "Periodic Condition 1" layer and the **(Right)** "Periodic Condition 2" layer of the parent "Electromagnetic Waves, Frequency Domain" layer. For both layers, "Floquet periodicity" is selected for the "Type of periodicity".

### S1.2.5: Mesh

The "Physics-controlled mesh" was chosen for "Sequence type:". The chosen "Element size:" was "Normal". Settings and mesh can be seen in Figure S11.

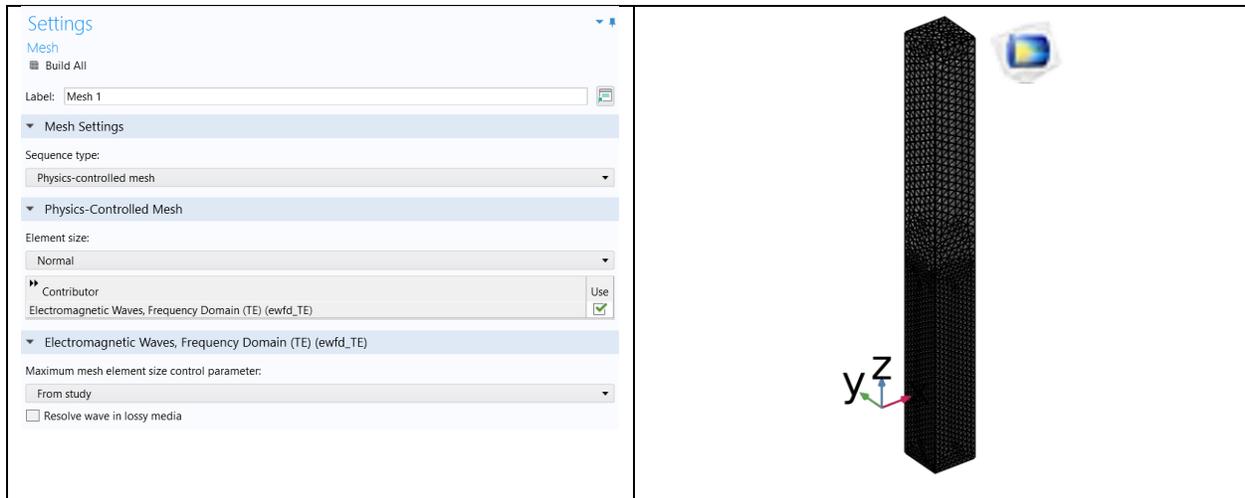

**Figure S11.** Settings of *(Left)* "Mesh" and *(Right)* the actual mesh of the model. The mesh layer is a child of "Component 1".

## Section S2: Predicting structural colour outputs

The reflectance spectra generated from section 1 were utilised to predict the structural colour output on the CIE 1931 2º Standard Observer colour space [6,7]. The XYZ tristimulus values quantify the amount of red, green, and blue recorded by the human eye and were calculated using the following equations:

$$X = k \sum_{\lambda} \phi_\lambda(\lambda)\bar{x}(\lambda)\Delta\lambda \tag{1}$$

$$Y = k \sum_{\lambda} \phi_\lambda(\lambda)\bar{y}(\lambda)\Delta\lambda \tag{2}$$

$$Z = k \sum_{\lambda} \phi_\lambda(\lambda)\bar{z}(\lambda)\Delta\lambda \tag{3}$$

The spectral distribution for a wavelength $\lambda$ was denoted by $\phi_\lambda(\lambda)$, [8], where $\lambda$ ranged between 400 nm and 700 nm. $\bar{x}$, $\bar{y}$, and $\bar{z}$ represent the colour matching functions of a standard colorimetric observer (Figure S12) [9]. $\Delta\lambda = 1\ nm$ denotes the step-size used for the wavelength measurements [8]. The constant $k$ was chosen such that $Y = 100$ for objects with reflectance equal to 1.0 across all wavelengths [8]:

$$k = \frac{100}{\sum_{\lambda} S(\lambda)\bar{y}(\lambda)\Delta\lambda} \tag{4}$$

$S(\lambda)$ is the spectral distribution of a selected illuminant [8]; in this work, the D65 illuminant was chosen [10,11] (Figure S13).

From these values, the $(x,y)$ chromaticity coordinates were derived [12] using the following equation:

$$x = \frac{X}{X + Y + Z} \tag{5}$$

$$y = \frac{Y}{X + Y + Z} \tag{6}$$

These coordinates enable a two-dimensional representation of the colour specified by the tristimulus values on the CIE 1931 2º Standard Observer colour space [12,13].

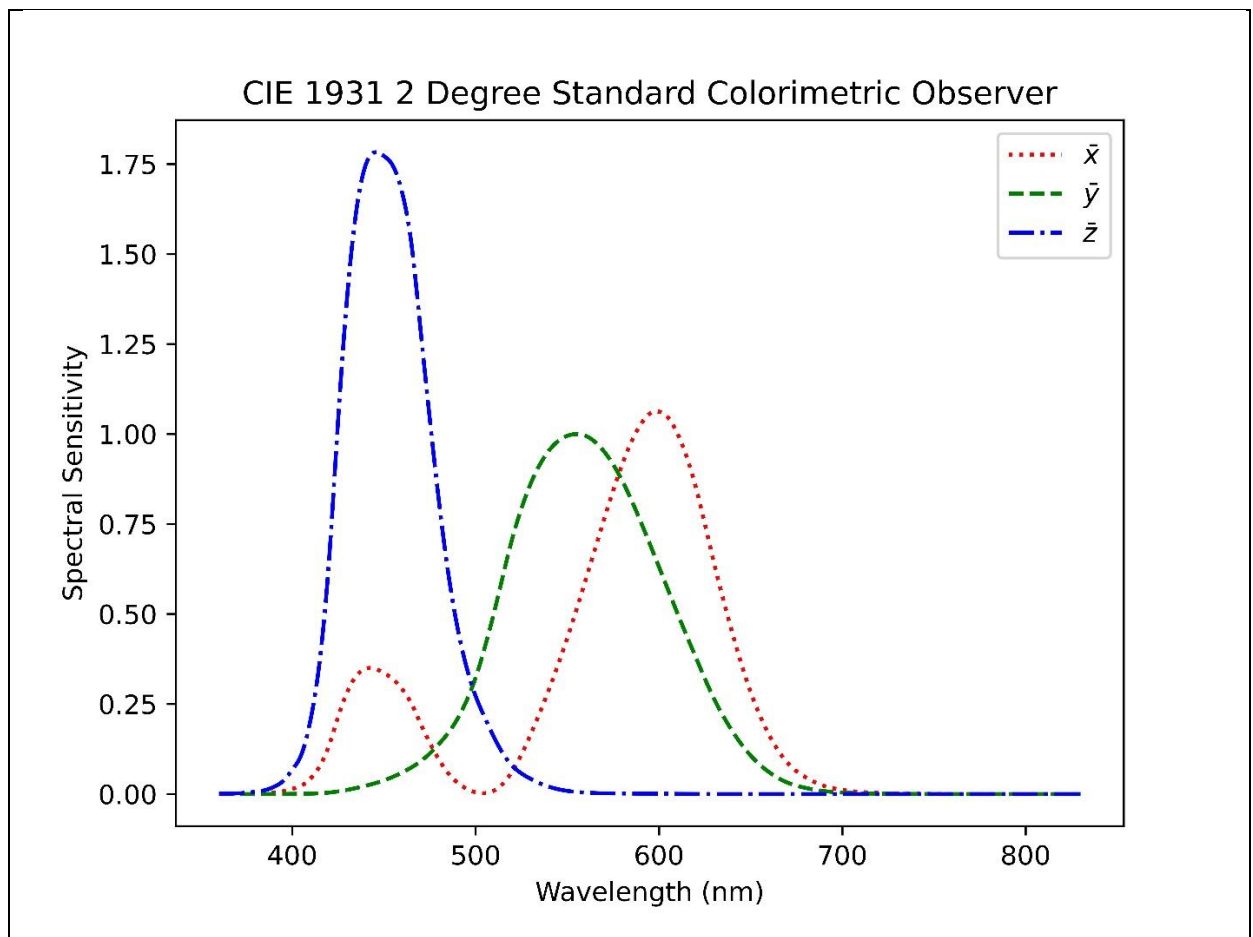

**Figure S12.** CIE 1931 2º degree standard colorimetric observer. Colour matching functions define how an observer's visual system responds to red, green, and blue primaries [6,7]. The functions were derived by interpolating empirical measurements and are used to calculate tristimulus values [6–8]. The colour matching functions utilised in this study were obtained from an open-source Python library called Colour (see Section S7.1 for programming details) [9]; and cross-checked with those provided by the International Commission on Illumination [6,10,11].

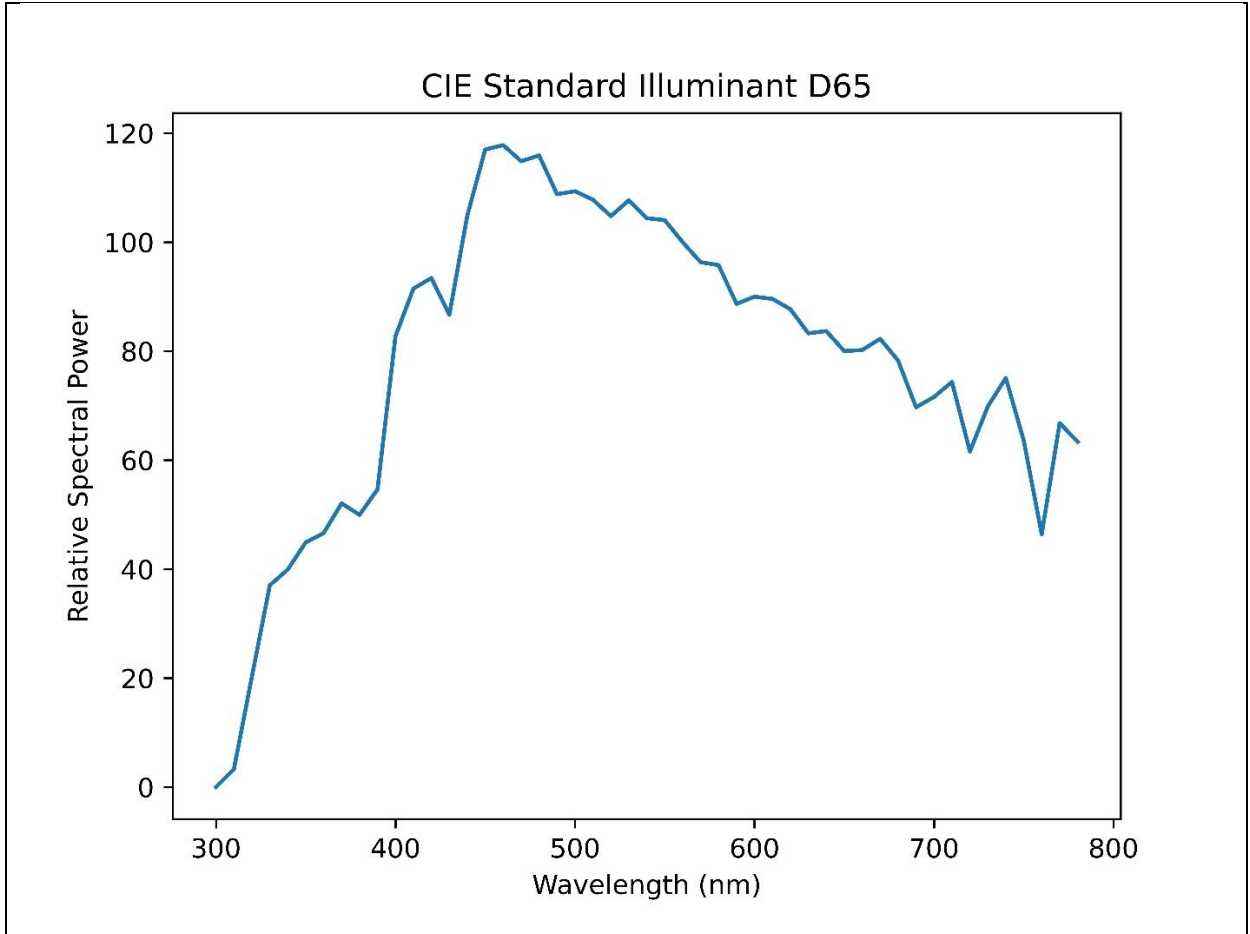

**Figure S13.** Relative spectral power distribution of the CIE standard illuminant D65. The CIE Standard Illuminant D65 represents average daylight in Northern and Western Europe, with a colour temperature of 6504 K [12,13]. Because it is an artificial illuminant, translation of our nanostructure arrays into the physical world is expected to yield a different colorimetric response under real light sources [14]. However, due to the challenges in obtaining tristimulus values for reflective media, such as our nanostructure arrays, we had to select a standard illuminant [14]. The choice of D65 is arbitrary. The illuminant spectrum was obtained from an open-source Python library called Colour (see Section S7.1 for programming details) [9]; and cross-checked with those provided by the International Commission on Illumination [15,16].

## Section S3: Evaluating colour saturation

The calculated XYZ tristimulus values from section 2 were converted to the L*a*b* colour space (Section S3.1) [14,15]. The resulting L*a*b* values were used to determine the colour saturation, $S_{LAB}$, as ratio of chroma, $C_{ab}^*$, to lightness, $L$ [16–19]:

$$S_{LAB} = \frac{C_{ab}^*}{L} = \frac{\sqrt{a^{*2} + b^{*2}}}{116 \left(\frac{Y}{Y_n}\right)^{\frac{1}{3}} - 16} \tag{7}$$

In the above equation, $a^*$ and $b^*$ correlate with red-green and blue-yellow chroma perceptions [15]. $Y_n$ is the tristimulus $Y$ value of the illuminant ($Y = 100$ for the D65 white point [20]) [11,15].

A qualitative comparison was conducted to assess the colour saturation of the colorimetric outputs of rectangle-shaped (or square-shaped) nanostructure arrays and their corresponding t-shaped nanostructure arrays. The authors are aware that Ref [19] finds discrepancies between this measure and the empirically-measured human perceptions of saturation. Nonetheless, saturation will be defined hereon as described above as it is the most accurate existing measure to the best of our knowledge.

## Section S3.1: Calculating L*a*b* values

The $L^*$, $a^*$, and $b^*$ quantities of the L*a*b* colour space correlate light-dark, red-green, and blue-yellow chroma perceptions, respectively. They are defined as follows [14,15]:

$$L^* = 116 \, f\left(\frac{Y}{Y_n}\right) - 16 \tag{8}$$

$$a^* = 500 \left[f\left(\frac{X}{X_n}\right) - f\left(\frac{Y}{Y_n}\right)\right] \tag{9}$$

$$b^* = 200 \left[f\left(\frac{Y}{Y_n}\right) - f\left(\frac{Z}{Z_n}\right)\right] \tag{10}$$

$$f(\omega) =$$
$$\begin{cases} \omega^{\frac{1}{3}}, & \omega > \left(\frac{24}{116}\right)^3 \\ \left(\frac{841}{108}\right)\omega + \frac{16}{116}, & w \leq \left(\frac{24}{116}\right)^3 \end{cases} \tag{11}$$

In equations 8-10, $X_n = 95.04$, $Y_n = 100.00$, and $Z_n = 108.88$ are the tristimulus values of the D65 white point [20].

**Section S4: Rectangle-shaped nanostructure arrays**

The length and width dimensions per aspect ratio of the rectangle-shaped nanostructure arrays (see Main Text, Results, Section 2) were chosen to keep a constant volume of 16,537,500 $nm^3$ [= 270 nm * length * width]. This was done to control for any effects the volume may have on the arising resonances, as per the advice of Ref. [21]. The exact geometric parameters are listed in Table S5.

**Table S5**. Length and width parameters for the simulated rectangle nanostructures seen in Section 2 of the Results in the Main Text.

| Aspect Ratio | 1:1 | 1:1.25 | 1:1.5 | 1:1.75 | 1:2 |
|---|---|---|---|---|---|
| Length (nm) | 247.5 | 276.7 | 303.1 | 327.4 | 350.0 |
| Width (nm) | 247.5 | 221.4 | 202.1 | 187.1 | 175.0 |
| Height (nm) | 270 | 270 | 270 | 270 | 270 |

### Section S5: Rectangle-bound t-shaped nanostructure arrays

Figure S15 shows the reflectance spectra of the 0-rad and π/2-rad linearly polarised light excitation of a rectangle-shaped nanostructure array (Figure S15A) and the corresponding t-shaped nanostructure arrays (Figures S15B-D), which corresponds to results reported in the Main Text.

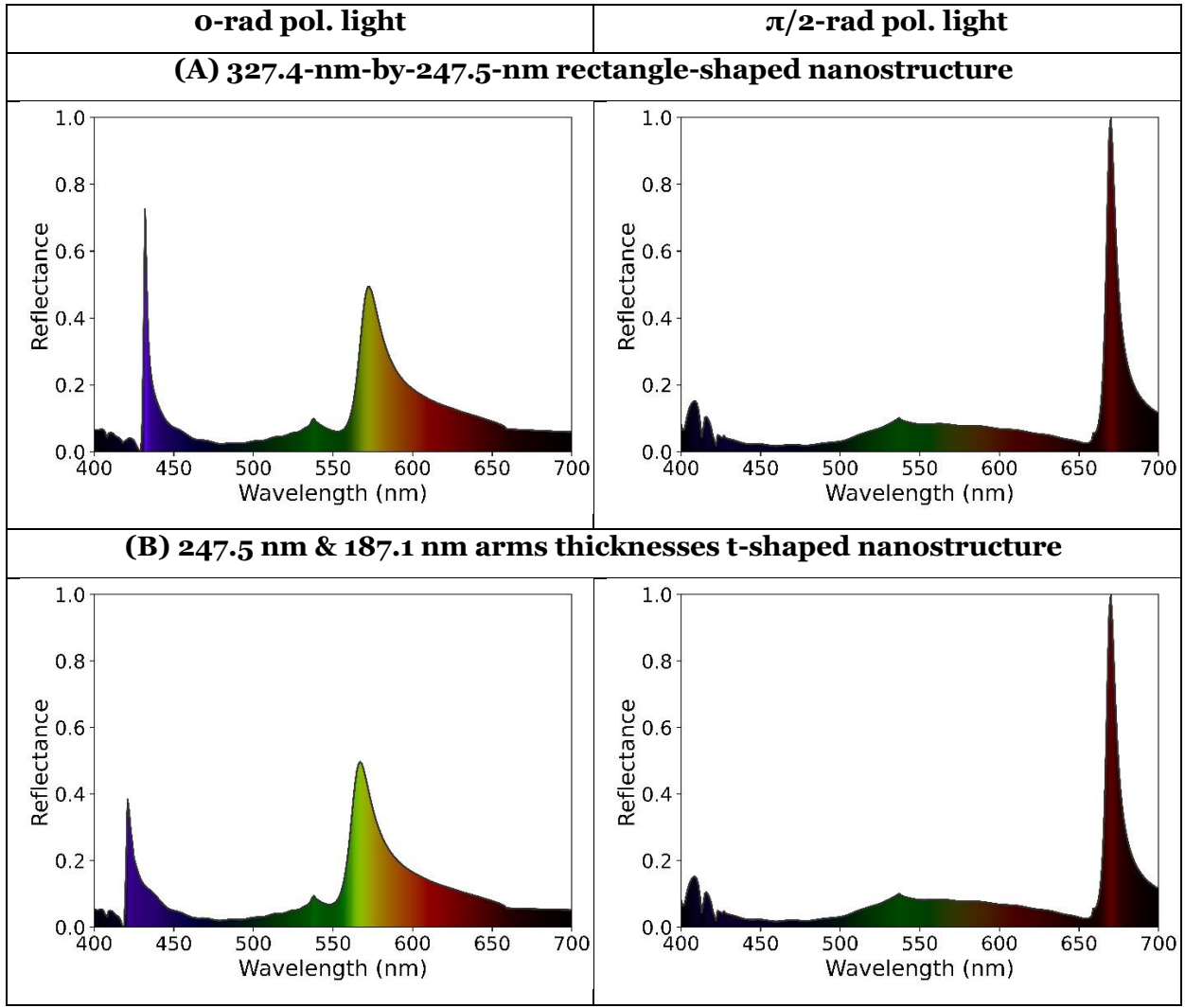

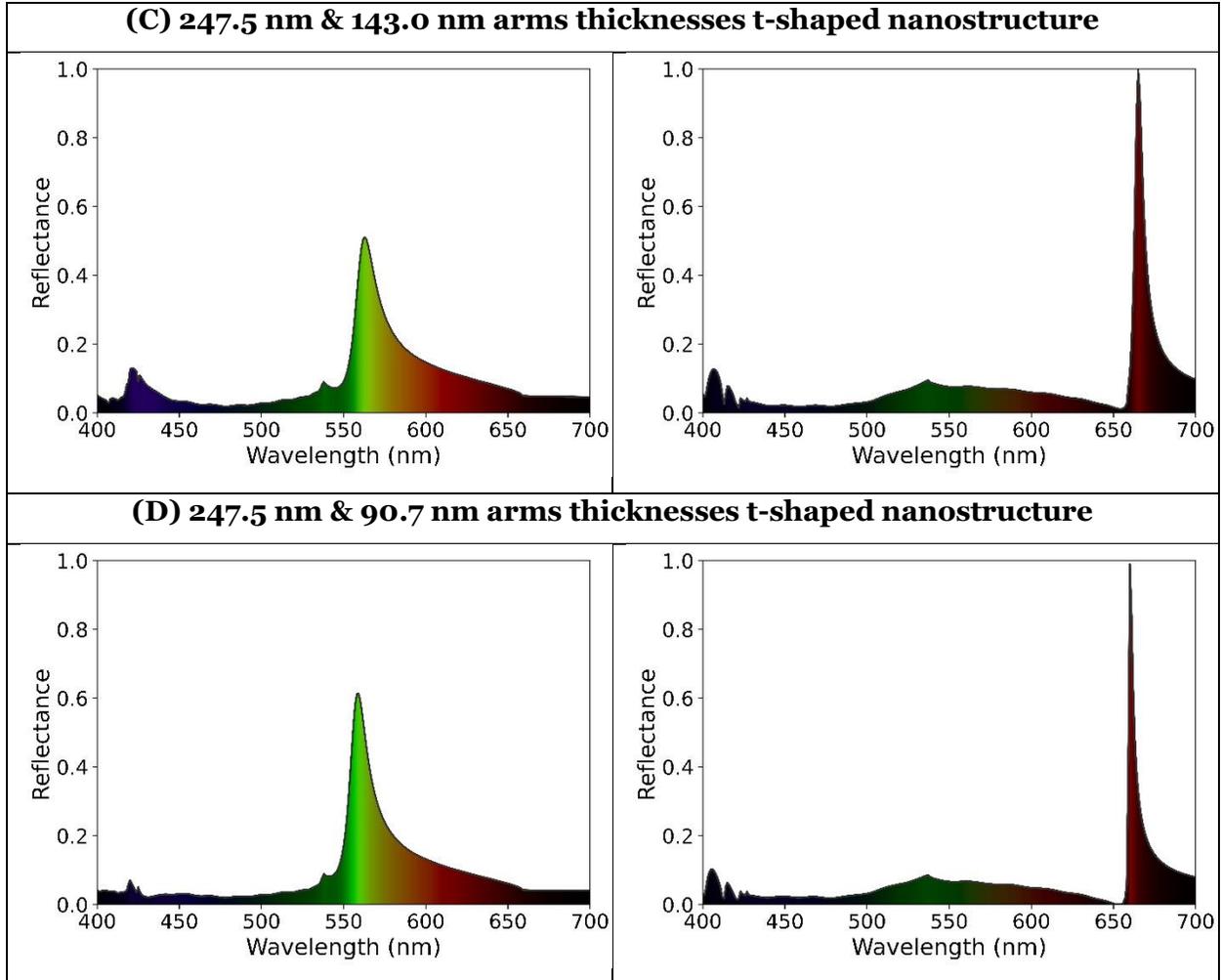

**Figure S15.** Reflectance plots of **(A)** rectangle-shaped and **(B-D)** t-shaped nanostructure arrays with 327.4-nm-by-247.5-nm bounds under (Left) 0-rad and (Right) π/2-rad linearly polarised light excitation. ***Left****: (A)* HRW: 432 nm, FRW: 572 nm, FWHM: 24 nm; *(B)* HRW: 421 nm, FRW: 567 nm, FWHM: 24 nm; *(C)* HRW: 421 nm, FRW: 563 nm, FWHM: 21 nm; *(D)* FRW: 559 nm, FWHM: 17 nm. ***Right****: (A)* FRW: 676 nm, FWHM: 8 nm; *(B)* FRW: 670 nm, FWHM: 8 nm; *(C)* FRW: 665 nm, FWHM: 7 nm; *(D)* FRW: 660 nm, FWHM: 4 nm. Abbreviations: FRW: Fundamental resonance wavelength; FWHM: Full width at half-maximum (of fundamental resonance); HRW: High-order resonance wavelength.

## Section S6: Square-bound t-shaped nanostructure arrays

### Section S6.1: Square-shaped nanostructure arrays

The main text highlights results from rectangle-shaped nanostructure arrays and their corresponding t-shape nanostructure arrays for the development of two-colour-bound sensors of incident polarised light. However, we also performed our experiments with square-shaped nanostructure arrays and their corresponding t-shape nanostructure arrays for the development of monochromatic pixel arrays. Figure S16 displays the setup, mechanism, and physics of square-shaped nanostructure arrays, which mimic those of the rectangle-shaped nanostructure arrays shown in the main text.

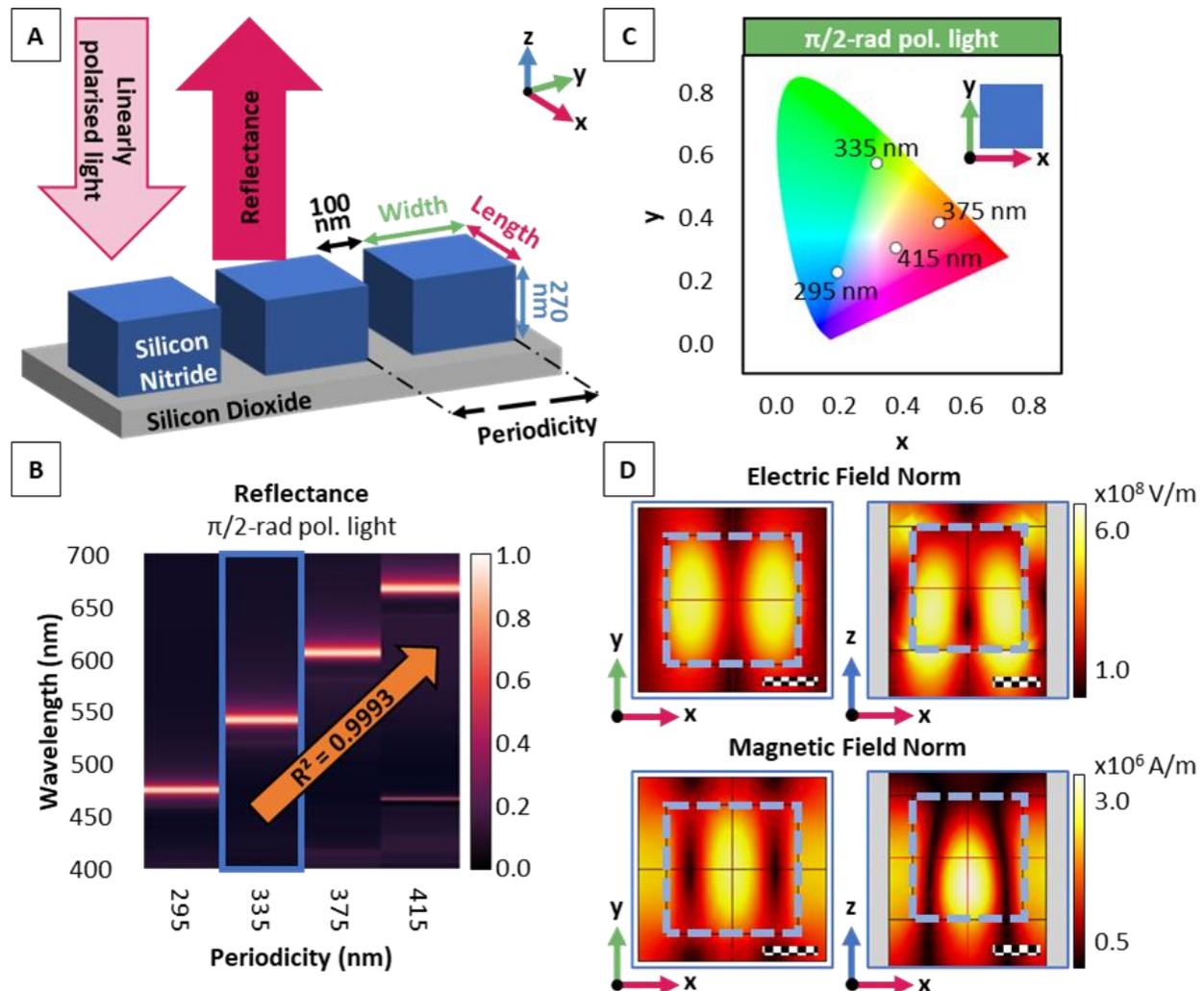

**Figure S16.** *(A)* Schematic of silicon nitride lattice resonant metasurface comprised of square-shaped nanostructures on a silicon dioxide substrate. *(B)* Heatmaps showing the reflectance spectra of square-shaped nanostructure arrays with periodicities of 295 nm, 335 nm, 375 nm, and 415 nm acquired by excitation with $\pi/2$ rad (x-axis) linearly polarised light. The 415-nm periodicity array displayed a high-order resonance at 466 nm. The orange arrow indicates shift of fundamental resonance wavelength with increasing periodicity. *(C)* Colorimetric output predicted from each reflectance spectrum plotted on the CIE 1931 2° Standard Observer colour space, illustrating the shift from blue to green to red as periodicity increases. *(D)* Near-field electric (top) and magnetic (bottom) norm plots of the 335-nm periodicity array. Checkered scale bars represent 100 nm.

**Section S6.2: Cutting corners of square-shaped nanostructures**

Using a 327.4-nm-by-327.4-nm square-shaped nanostructure array as a control, two resonance peaks were observed under excitation by linearly polarised light at 0 rad (Figure S17A): a high-order resonance at 481 nm with amplitude 0.89 and a fundamental resonance at 685 nm with amplitude 1.0. By cutting the corners of the nanostructures, a t-shaped nanostructure array with nanostructure arm thicknesses of 234.4 nm along both the vertical and horizontal axes was created (Figure S17B). This modification resulted in a blue-shift of the high-order and fundamental resonance peaks to 469 nm with amplitude 0.37 and 676 nm with amplitude 0.99, respectively. A second corner cut of the t-shape to arm thicknesses of 162.6 nm and 234.4 nm along the vertical and horizontal axes (Figure S17C) maintained the high-order resonance peak at 469 nm with amplitude 0.14 and the fundamental resonance peak at 668 nm with amplitude 1.0. A third corner cut of the t-shape to arm thicknesses of 162.6 nm along both the vertical and horizontal axes (Figure S17D) extinguished the high-order resonance and blue-shifted the fundamental resonance peak to 660 nm with amplitude 0.50.

Examining the electric field norms of the four nanostructure arrays at the high-order resonance peaks (Figures S17E-H, Left), revealed a decrease in the maximum value from 1.03E9 V/m with the control structure to 6.94E8 V/m, 3.61E8 V/m, and 1.7E8 with the first, second, and third corner-cut structures, respectively. In contrast, the electric field norms of the four nanostructure arrays at the fundamental resonance peaks (Figure S17E-H, Right) showed a maximum value of 4.81E8 V/m with the control structure and maximum values of 5.4E8 V/m, 5.77E8 V/m, and 5.31E8 V/m with the first, second, and third corner-cut structures, respectively. The high-order resonances exhibited activities centred at the corners and centres of the nanostructures, while the fundamental resonances showed activity only in the centres of the nanostructures.

Exciting the control and first corner-cut nanostructure arrays with linearly polarised light at angles of 0 rad, $\pi/6$ rad, $\pi/4$ rad, $\pi/3$ rad, and $\pi/2$ rad resulted in identical reflectance spectra across all polarisations (Figures S18A and S18B). The colorimetric output of these spectra on the CIE 1931 2° Standard Observer colour space appeared as a single point (Figures S18C and S18D).

In contrast, exciting the second corner-cut nanostructure array, where the corners were asymmetrically cut, with linearly polarised light resulted in hybrid signals comprising resonances from the 0 rad and $\pi/2$ rad polarisations (Figure S18E). The 0 rad polarisation displayed a high-order resonance whereas the $\pi/2$ rad polarisation did not. These hybrid signals were amplitude-adjusted based on the angular deviation from the 0 rad or $\pi/2$ rad axes. The colorimetric output of these spectra on the CIE 1931 2º Standard Observer colour space appeared on a line bounded between the colorimetric outputs of the 0 rad and π/2 rad polarisations (Figure S18F).

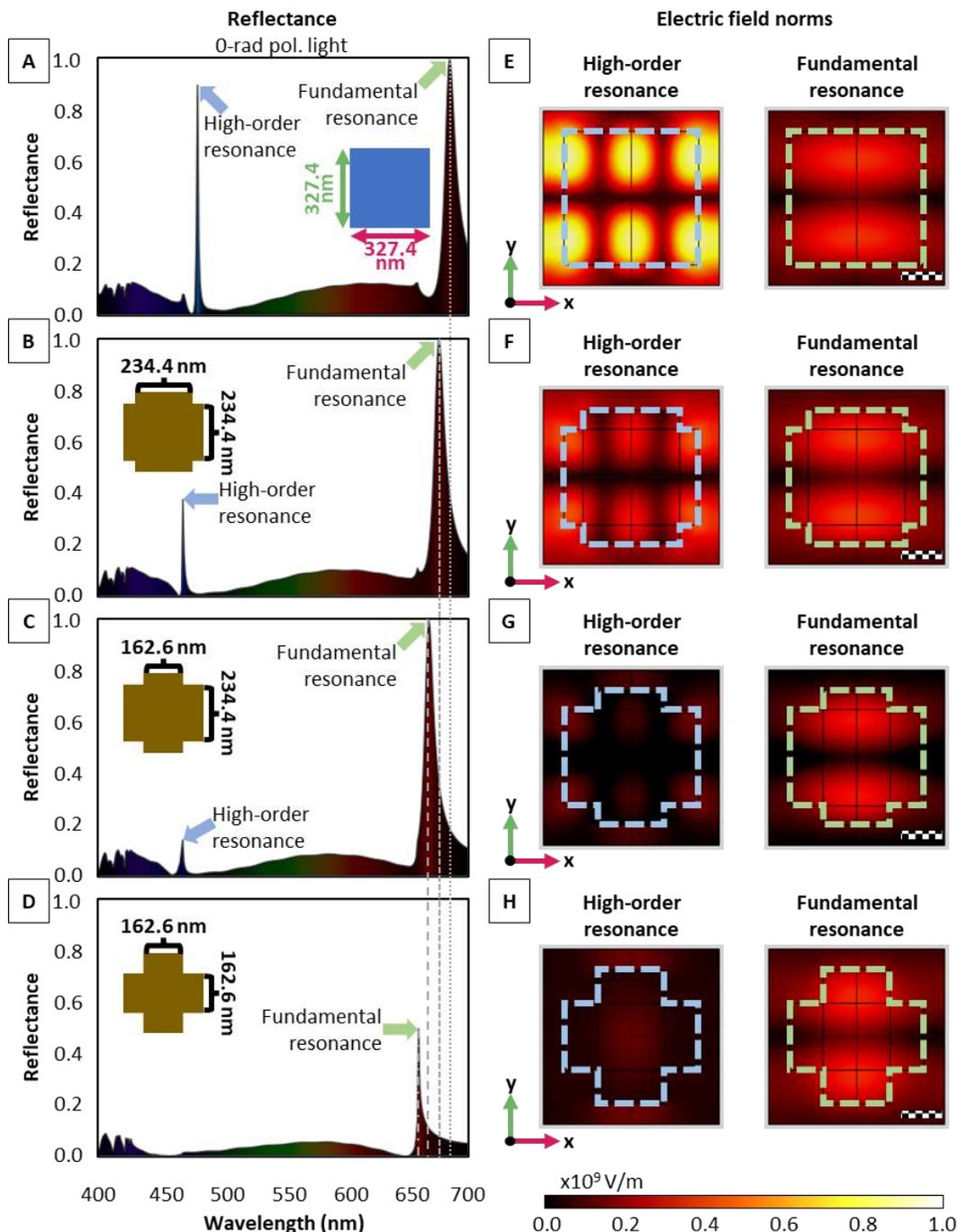

**Figure S17.** *(A-D)* Reflectance spectra of square-shaped and t-shaped nanostructure arrays, indicated by insets, under 0 rad (y-axis) linearly polarised light excitation. Green arrows highlight the location of the fundamental resonance wavelength. Blue arrows highlight the location of the high-order resonance wavelength. The dotted, short dashed, long dashed, and dash-dotted lines indicate the position of the fundamental resonance wavelength, which is blue-shifted when transforming a square-shaped nanostructure array to a t-shaped nanostructure array. *(E-H)* Near-field plots of the electric field norm of the (Left) high-order and (Right) fundamental resonances of the square-shaped and t-shaped nanostructure arrays. Checkered scale bars are 100 nm.

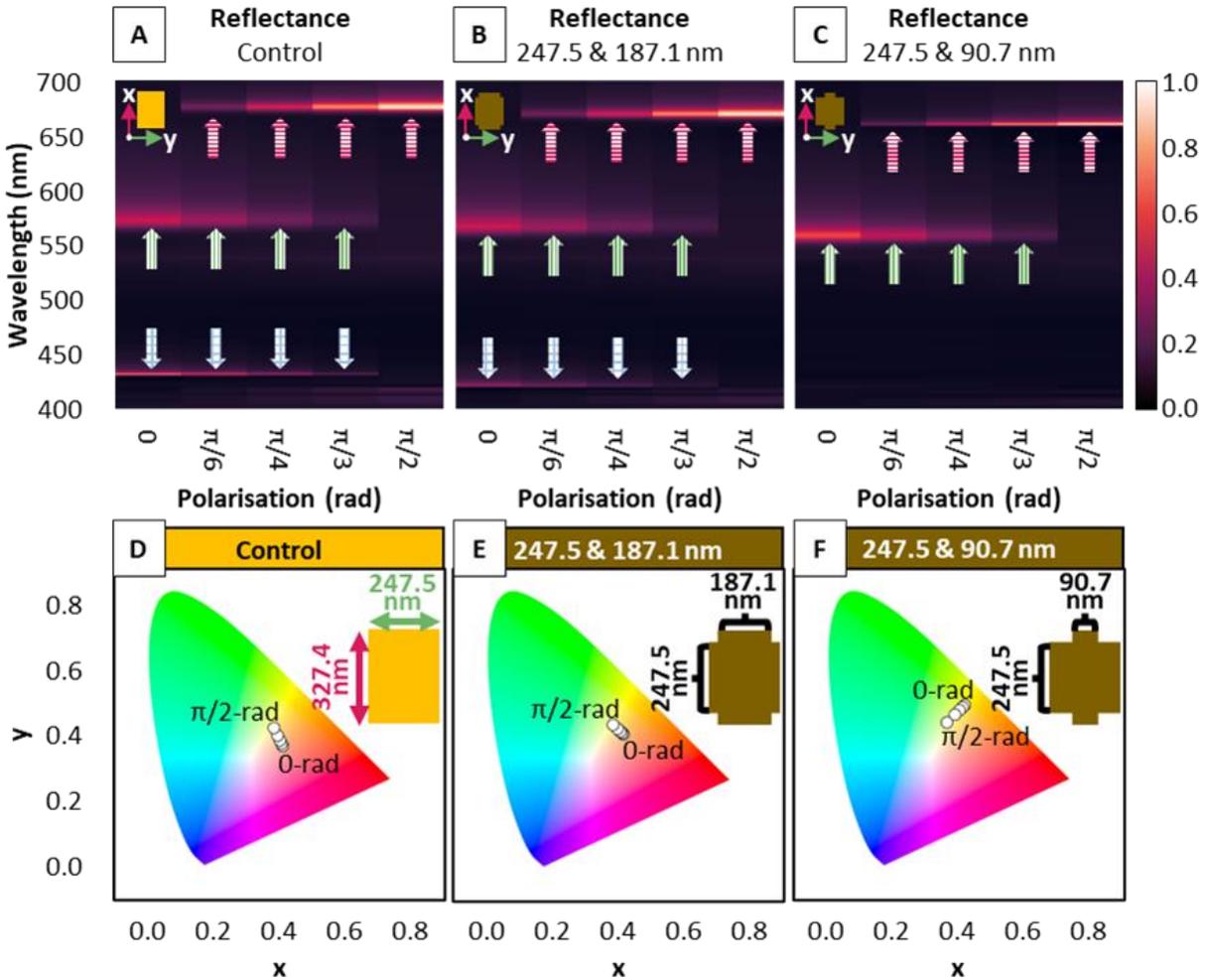

**Figure S18.** *(A-C)* Reflectance heatmaps acquired from the rectangle-shaped and t-shaped nanostructure arrays from Figure 3. Insets indicate the type of nanostructure. Blue, grid-filled arrows indicate the high-order resonances. Green, vertically striped arrows indicate fundamental resonance signals shared with the 0 rad (y-axis) polarisation. Red, horizontally striped arrows indicate fundamental resonance signals shared with the π/2 rad (x-axis) polarisation. The high-order resonances vanished with the t-shaped nanostructure array in (C). *(D-F)* Colorimetric output predicted from each reflectance spectrum in (A-C) plotted on the CIE 1931 2° Standard Observer colour space. Only the 0 rad (y-axis) and π/2 rad (x-axis) polarisations are labelled. All outputs are bounded along a line. The lines in (D, F) have a diagonal orientation. The line in (F) has an anti-diagonal orientation.

## Section S6.4: Saturation scores of square-bound nanostructure arrays

The colour saturation was calculated as described in Section S3 of this document and shown in Figure S19. In the square-bound t-shaped structures, both the 0 rad polarisation and π/2 rad polarisation exhibited high-order and fundamental resonances (see Supplementary Material, Section S6.5 for corresponding reflectance spectra). Under 0 rad polarisation, the colour saturation decreased as the area of the removed corners increased. Under π/2 rad polarisation, the colour saturation only showed an increase when the largest area of the corners was removed, while the other t-shaped structures showed a decrease in colour saturation compared to the control. This decrease in colour saturation is attributed to (i) the reduced capability of these nanostructures to support the fundamental resonance across a wider range of wavelengths, evidenced by a decrease in the full-width at half maximum of these resonances (see caption in Supplementary Material, Figure S20), and (ii) the failure to extinguish the high-order resonances (Figures S17A-C). Figure S17D showed the dampening consequence of cutting the corners too close to the activity region of the fundamental resonance.

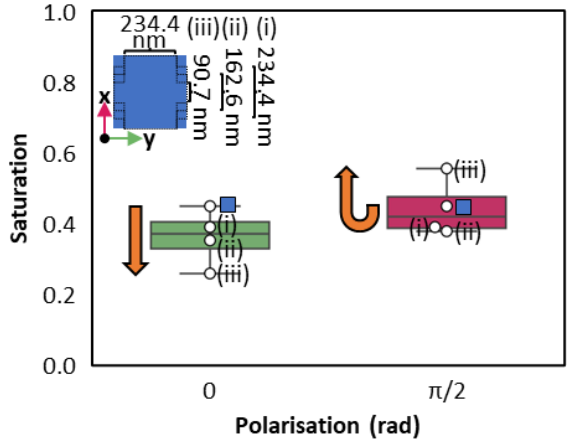

**Figure S19.** Box-and-whisker plot illustrating colour saturation measurements of square-bound nanostructure arrays under 0 rad (y-axis) and π/2 rad (x-axis) linearly polarised light excitation. (i), (ii), and (iii) indicate t-shape structures where (i) has the smallest area of corners removed and (iii) with the largest area of corners removed (exact dimensions provided in insets). Orange arrows indicate movement from the square-shaped control to the t-shape with the greatest area of corners cut out. Boxes represent the quartiles, with the horizontal grey line indicating the median. Whiskers represent the range of the distribution.

**Section S6.5: Orthogonal polarisations reflectance spectra**

      Figure S19 shows the reflectance spectra of the 0-rad and π/2-rad linearly polarised light excitation of a square-shaped nanostructure array (Figure S16A) and the corresponding t-shaped nanostructure arrays (Figures S16B-D). These results are not reported in the main text.

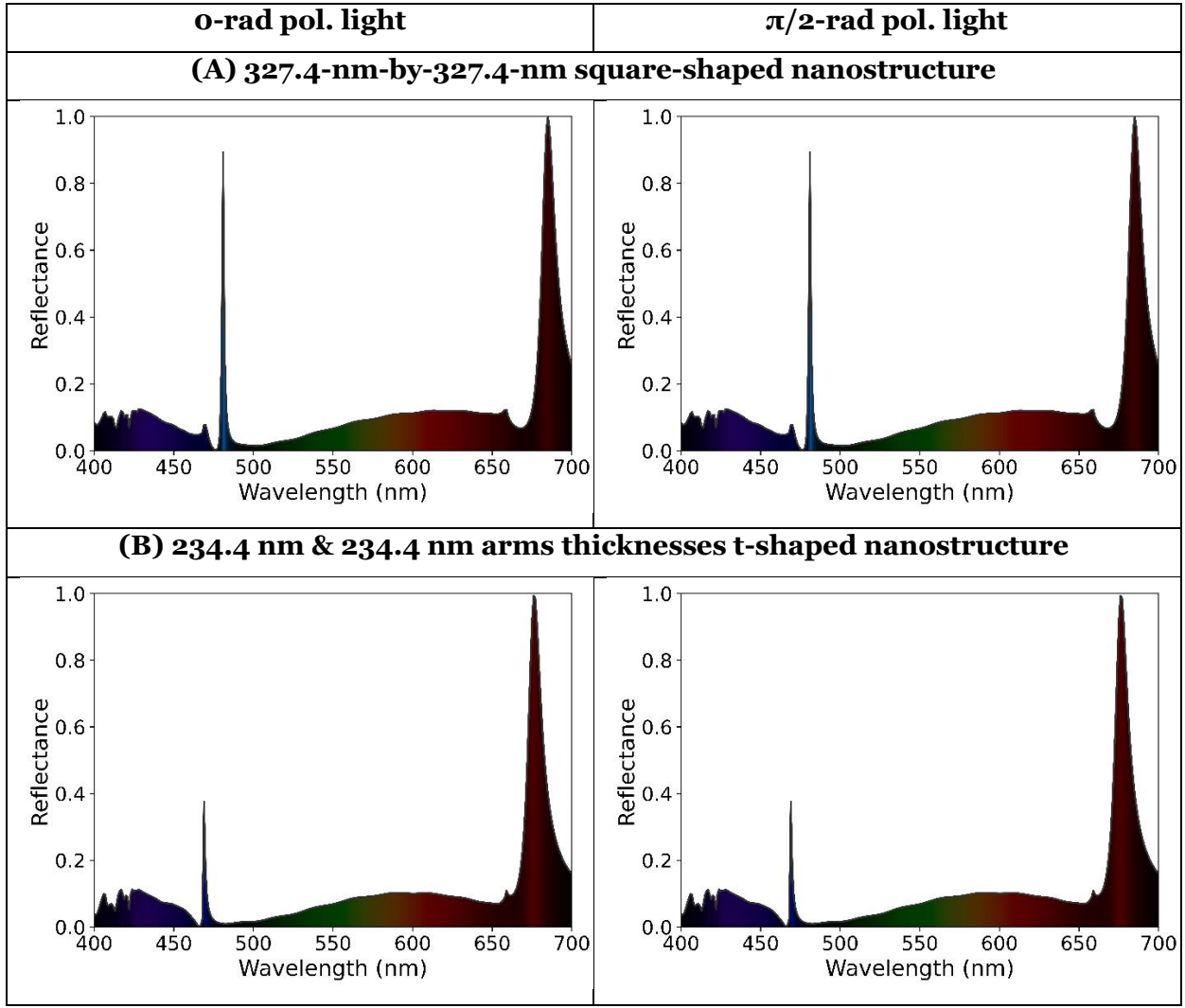

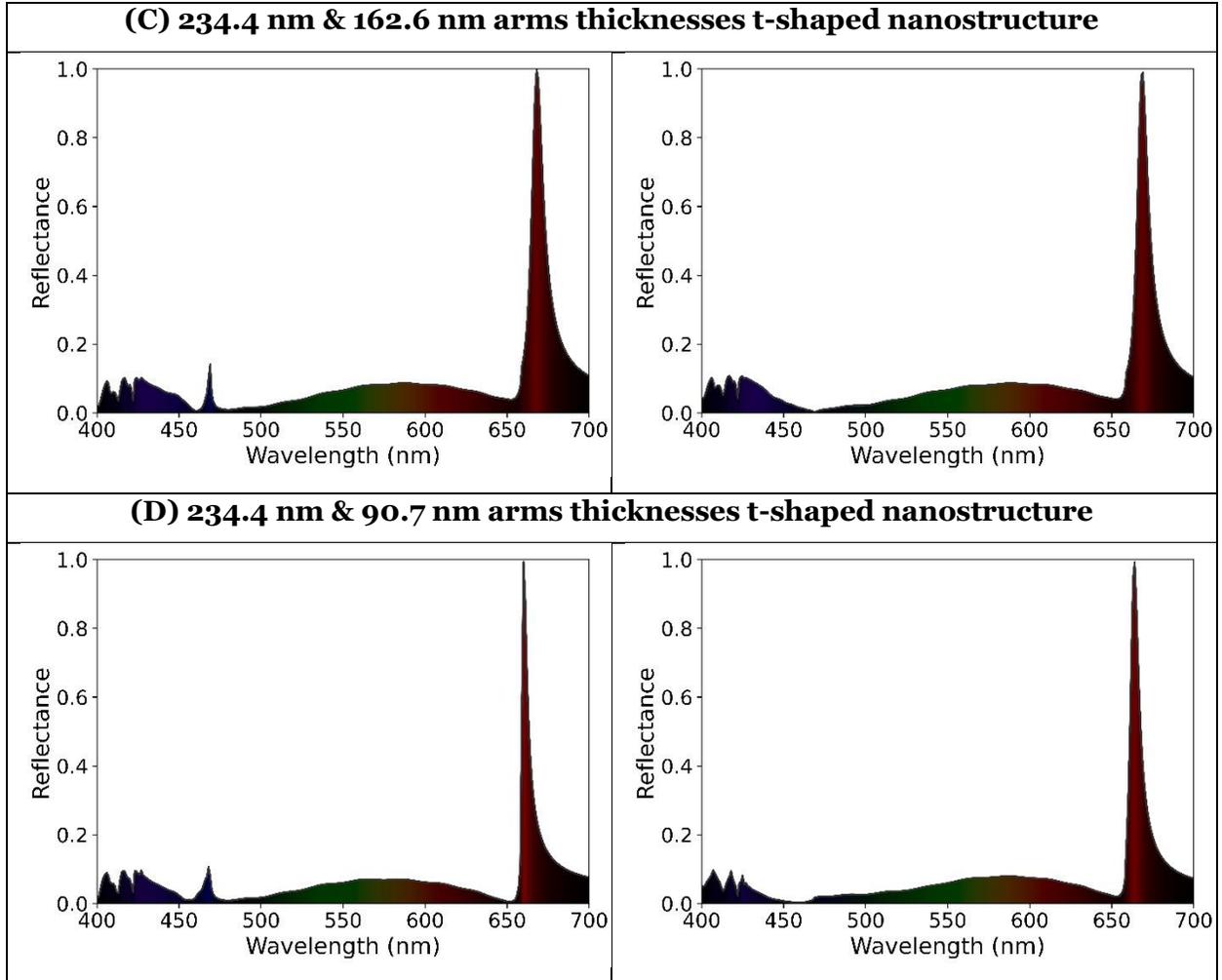

**Fig. S20:** Reflectance plots of *(A)* square-shaped and *(B-D)* t-shaped nanostructure arrays with 327.4-nm-by-327.4-nm bounds under (Left) 0-rad and (Right) π/2-rad linearly polarised light excitation. ***Left:*** *(A)* HRW: 481 nm, FRW: 685 nm, FWHM: 11 nm; *(B)* HRW: 469 nm, FRW: 676 nm, FWHM: 11 nm; *(C)* HRW: 469 nm, FRW: 668 nm, FWHM: 10 nm; *(D)* HRW: 468 nm, FRW: 660 nm, FWHM: 4 nm. ***Right:*** *(A)* HRW: 481 nm, FRW: 685 nm, FWHM: 11 nm; *(B)* HRW: 469 nm, FRW: 676 nm, FWHM: 11 nm; *(C)* FRW: 669 nm, FWHM: 9 nm; *(D)* FRW: 664 nm, FWHM: 7 nm. Abbreviations: FRW: Fundamental resonance wavelength; FWHM: Full width at half-maximum (of fundamental resonance); HRW: High-order resonance wavelength.

## Section S7: Programming Scripts

All programming scripts and environments are provided in the "Workflow/Programming/" folder. Within this folder are the "Python" (Section S7.1) and "Mathematica" (Section S7.2) sub-folders.

### Section S7.1: Python

Conda v22.11.1 [22] installed using the Miniconda package [23] by Anaconda Software Distribution [24] was utilised to create a virtual programming environment. Within this environment, Python v3.10.8 [25,26] was installed. This Python version, in conjunction with several open-source libraries, was used for the calculations and plots presented in the Main Text. The primary libraries used were Colour v0.4.2 [27], Matplotlib v3.6.2 [28,29], Numpy v1.23.5 [30,31], Pandas v1.5.2 [32,33], and Seaborn v0.12.2 [34]. A full list of packages and dependencies installed within the Conda environment can be found in Section S7.1.2.

The Conda environment file is provided as "20230718_CuttingCorners_CondaEnvironment.yml". Instructions for activating this environment can be found in Ref. [35].

In addition, three Python files are provided:

1. "Main.py": Central hub for analysing and plotting the generated data.
2. "SpectralAnalysis.py": Implements a Spectrum class to keep track of a given spectrum's metadata and resultant calculations.
3. "ColourPlotTemplates.py": Provides functions for visualising the data extracted from a spectrum or spectra.'

All programming for this section was completed on Ubuntu 22.04.1 LTS [36] installed in the WLS2 environment of Windows 10 [37].

### *Section S7.1.2: Python Packages & Dependencies*

```
# packages in environment at /home/███████/.conda/envs/DataScience_env:
#
# Name                    Version                   Build  Channel
_libgcc_mutex             0.1                        main
_openmp_mutex             5.1                       1_gnu
astroid                   2.11.7          py310h06a4308_0    anaconda
asttokens                 2.2.1              pyhd8ed1ab_0    conda-forge
backcall                  0.2.0              pyh9f0ad1d_0    conda-forge
backports                 1.0                pyhd8ed1ab_3    conda-forge
backports.functools_lru_cache 1.6.4          pyhd8ed1ab_0    conda-forge
beartype                  0.11.0             pyhd8ed1ab_0    conda-forge
blas                      1.0                         mkl
bottleneck                1.3.5           py310ha9d4c09_0
brotli                    1.0.9               h5eee18b_7
brotli-bin                1.0.9               h5eee18b_7
bzip2                     1.0.8               h7b6447c_0
ca-certificates           2022.12.7           ha878542_0    conda-forge
```

```
certifi                 2022.12.7       pyhd8ed1ab_0      conda-forge
colour-science          0.4.2           pyh6c4a22f_0      conda-forge
contourpy               1.0.5           py310hdb19cb5_0
cycler                  0.11.0          pyhd3eb1b0_0
dbus                    1.13.18         hb2f20db_0
debugpy                 1.5.1           py310h295c915_0
decorator               5.1.1           pyhd8ed1ab_0      conda-forge
dill                    0.3.4           pyhd3eb1b0_0      anaconda
executing               1.2.0           pyhd8ed1ab_0      conda-forge
expat                   2.4.9           h6a678d5_0
fftw                    3.3.9           h27cfd23_1
fontconfig              2.14.1          h52c9d5c_1
fonttools               4.25.0          pyhd3eb1b0_0
freetype                2.12.1          h4a9f257_0
geos                    3.8.0           he6710b0_0
giflib                  5.2.1           h7b6447c_0
glib                    2.69.1          he621ea3_2
gst-plugins-base        1.14.0          h8213a91_2
gstreamer               1.14.0          h28cd5cc_2
icu                     58.2            he6710b0_3
imageio                 2.25.0          pyh24c5eb1_0      conda-forge
importlib-metadata      6.0.0           pyha770c72_0      conda-forge
importlib_metadata      6.0.0           hd8ed1ab_0        conda-forge
intel-openmp            2021.4.0        h06a4308_3561
ipykernel               6.15.0          pyh210e3f2_0      conda-forge
ipython                 8.10.0          pyh41d4057_0      conda-forge
isort                   5.9.3           pyhd3eb1b0_0      anaconda
jedi                    0.18.2          pyhd8ed1ab_0      conda-forge
joblib                  1.1.1           py310h06a4308_0
jpeg                    9e              h7f8727e_0
jupyter_client          8.0.3           pyhd8ed1ab_0      conda-forge
jupyter_core            4.12.0          py310hff52083_0   conda-forge
kiwisolver              1.4.4           py310h6a678d5_0
krb5                    1.19.2          hac12032_0
lazy-object-proxy       1.6.0           py310h7f8727e_0   anaconda
lcms2                   2.12            h3be6417_0
ld_impl_linux-64        2.38            h1181459_1
lerc                    3.0             h295c915_0
libbrotlicommon         1.0.9           h5eee18b_7
libbrotlidec            1.0.9           h5eee18b_7
libbrotlienc            1.0.9           h5eee18b_7
libclang                10.0.1          default_hb85057a_2
libdeflate              1.8             h7f8727e_5
libedit                 3.1.20221030    h5eee18b_0
libevent                2.1.12          h8f2d780_0
libffi                  3.4.2           h6a678d5_6
libgcc-ng               11.2.0          h1234567_1
libgfortran-ng          11.2.0          h00389a5_1
libgfortran5            11.2.0          h1234567_1
libgomp                 11.2.0          h1234567_1
libllvm10               10.0.1          hbcb73fb_5
libpng                  1.6.37          hbc83047_0
libpq                   12.9            h16c4e8d_3
libsodium               1.0.18          h36c2ea0_1        conda-forge
libstdcxx-ng            11.2.0          h1234567_1
libtiff                 4.5.0           hecacb30_0
libuuid                 1.41.5          h5eee18b_0
```

```
libwebp                 1.2.4            h11a3e52_0
libwebp-base            1.2.4            h5eee18b_0
libxcb                  1.15             h7f8727e_0
libxkbcommon            1.0.1            hfa300c1_0
libxml2                 2.9.14           h74e7548_0
libxslt                 1.1.35           h4e12654_0
lz4-c                   1.9.4            h6a678d5_0
matplotlib              3.6.2            py310h06a4308_0
matplotlib-base         3.6.2            py310h945d387_0
matplotlib-inline       0.1.6            pyhd8ed1ab_0      conda-forge
mccabe                  0.7.0            pyhd3eb1b0_0      anaconda
mkl                     2021.4.0         h06a4308_640
mkl-service             2.4.0            py310h7f8727e_0
mkl_fft                 1.3.1            py310hd6ae3a3_0
mkl_random              1.2.2            py310h00e6091_0
munkres                 1.1.4            py_0
ncurses                 6.3              h5eee18b_3
nest-asyncio            1.5.6            pyhd8ed1ab_0      conda-forge
nspr                    4.33             h295c915_0
nss                     3.74             h0370c37_0
numexpr                 2.8.4            py310h8879344_0
numpy                   1.23.5           py310hd5efca6_0
numpy-base              1.23.5           py310h8e6c178_0
openssl                 1.1.1t           h7f8727e_0
packaging               22.0             py310h06a4308_0
pandas                  1.5.2            py310h1128e8f_0
parso                   0.8.3            pyhd8ed1ab_0      conda-forge
pcre                    8.45             h295c915_0
pep8                    1.7.1            py310h06a4308_1   anaconda
pexpect                 4.8.0            pyh1a96a4e_2      conda-forge
pickleshare             0.7.5            py_1003           conda-forge
pillow                  9.3.0            py310hace64e9_1
pip                     22.3.1           py310h06a4308_0
platformdirs            2.4.0            pyhd3eb1b0_0      anaconda
ply                     3.11             py310h06a4308_0
prompt-toolkit          3.0.36           pyha770c72_0      conda-forge
psutil                  5.9.0            py310h5eee18b_0
ptyprocess              0.7.0            pyhd3deb0d_0      conda-forge
pure_eval               0.2.2            pyhd8ed1ab_0      conda-forge
pygments                2.14.0           pyhd8ed1ab_0      conda-forge
pylint                  2.14.5           py310h06a4308_0   anaconda
pyparsing               3.0.9            py310h06a4308_0
pyqt                    5.15.7           py310h6a678d5_1
pyqt5-sip               12.11.0          pypi_0            pypi
python                  3.10.8           h7a1cb2a_1
python-dateutil         2.8.2            pyhd3eb1b0_0
python_abi              3.10             2_cp310           conda-forge
pytz                    2022.7           py310h06a4308_0
pyzmq                   23.2.0           py310h6a678d5_0
qt-main                 5.15.2           h327a75a_7
qt-webengine            5.15.9           hd2b0992_4
qtwebkit                5.212            h4eab89a_4
readline                8.2              h5eee18b_0
scikit-learn            1.2.0            py310h6a678d5_0
scipy                   1.9.3            py310hd5efca6_0
seaborn                 0.12.2           py310h06a4308_0
setuptools              65.6.3           py310h06a4308_0
```

```
shapely                 1.8.4           py310h81ba7c5_0
sip                     6.6.2           py310h6a678d5_0
six                     1.16.0            pyhd3eb1b0_1
sqlite                  3.40.1             h5082296_0
stack_data              0.6.2            pyhd8ed1ab_0      conda-forge
threadpoolctl           2.2.0            pyh0d69192_0
tk                      8.6.12            h1ccaba5_0
toml                    0.10.2           pyhd3eb1b0_0
tomli                   2.0.1           py310h06a4308_0    anaconda
tomlkit                 0.11.1          py310h06a4308_0    anaconda
tornado                 6.2             py310h5eee18b_0
traitlets               5.9.0            pyhd8ed1ab_0      conda-forge
typing_extensions       4.3.0           py310h06a4308_0    anaconda
tzdata                  2022g             h04d1e81_0
wcwidth                 0.2.6            pyhd8ed1ab_0      conda-forge
wheel                   0.37.1           pyhd3eb1b0_0
wrapt                   1.14.1          py310h5eee18b_0    anaconda
xz                      5.2.8             h5eee18b_0
zeromq                  4.3.4             h9c3ff4c_1       conda-forge
zipp                    3.14.0           pyhd8ed1ab_0      conda-forge
zlib                    1.2.13            h5eee18b_0
zstd                    1.5.2             ha4553b6_0
```

## Section S7.2: Mathematica

Wolfram notebook (extension: .nb) was used with  Wolfram Mathematica v13.3.0.0 [38,39] for 64-bit Microsoft Windows 11. Colour matching functions and standard illuminant D65 values were imported from the online datasets provided by the International Commission on Illumination [40,41]. The analyses performed with Mathematica were not included in the Main Text; it was used as a qualitative validation of the analysis performed in Python (see Section S8.1).

Nevertheless, the Mathematica code is provided in a file named, "SpectralAnalysis.nb". Further details of the code's implementation are provided in Section S7.2.1.

### Section S7.2.1: Code

The code within "SpectralAnalysis.nb" can be split into four sections (labelled within the code):

(i) "Import XYZ Functions and D65 White point": imports and plots the colour matching functions and standard illuminant D65 values; and calculates the D65 tristimulus values normalised such that Y=1. To note, the tristimulus value calculations here do not consider the normalisation factor mentioned in Section 2 of the Main Text. Thus, values acquired from Mathematica can only be compared qualitatively and not quantitatively with those from Python.

(ii) "Define XYZ -> L*a*b* conversion with D65 reference white": Converts the D65 tristimulus values from (i) to their corresponding $L^*$, $a^*$, and $b^*$ quantities (see Main Text, Appendix, Section A1, Equations 8-11).

(iii) "COMSOL Data Import": Imports the spectra data collected from the COMSOL Multiphysics® simulations.

(iv) "Calculate LAB co-ordinates of Spectra, their Distance to D65 and Saturation": Calculates the $L^*$, $a^*$, and $b^*$ quantities of the imported spectra and their corresponding saturation scores (see Main Text, Methods, Section 3, Equation 7). These results were exported out as .csv tables.

To note, this code provides additional data calculations and visualisations not described above. However, what is described above is predominantly what was used for the qualitative comparisons with the data processed in Python that was reported in the Main Text.

## Section S8: Data

All data, found in the "Workflow/Data/" folder, is provided in either .txt or .csv format. Columns in .txt files are separated with space(s); columns in .csv files are separated by a comma (","). Each file contains four columns and a corresponding header. The attributes of each column are as follows:

1. Column #1: Wavelength(nm)
2. Column #2: Reflectance
3. Column #3: Transmittance
4. Column #4: Absorptance

There are a total of fifty-three files. The files are organised into appropriate folders titled "SquareShapedNanostructureArrays" (Section S8.1), "RectangleShapedNanostructureArrays" (Section S8.2), "RectangleBoundTShapedNanostructureArrays" (SectionS8.3), and "SquareBoundTShapedNanostructureArrays" (Section S8.4).

### Section S8.1: Square-shaped nanostructure arrays

Four .csv files are provided containing reflectance, transmittance, and absorptance data of square-shaped nanostructure arrays excited by $\pi/2$-rad linearly polarised light:

1. P195nm_SquareNanoarray_Study01_RTATable.csv
2. P235nm_SquareNanoarray_Study01_RTATable.csv
3. P275nm_SquareNanoarray_Study01_RTATable.csv
4. P315nm_SquareNanoarray_Study01_RTATable.csv

The number following the "P" flag in each file name corresponds to the periodicity of the array: 195 nm, 235 nm, 275 nm, or 315 nm.

### Section S8.2: Rectangle-shaped nanostructure arrays

Thirteen .csv files are provided containing reflectance, transmittance, and absorptance data of rectangle-shaped nanostructure arrays excited by varying angles of linearly polarised light:

1. 0degpol_RectangleNanoarray_Study00_RTATable.txt
2. 0degpol_RectangleNanoarray_Study01_RTATable.txt
3. 0degpol_RectangleNanoarray_Study02_RTATable.txt
4. 0degpol_RectangleNanoarray_Study03_RTATable.txt
5. 0degpol_RectangleNanoarray_Study04_RTATable.txt
6. 30degpol_RectangleNanoarray_Study01_RTATable.txt
7. 45degpol_RectangleNanoarray_Study01_RTATable.txt
8. 60degpol_RectangleNanoarray_Study01_RTATable.txt
9. 90degpol_RectangleNanoarray_Study00_RTATable.txt

10. 90degpol_RectangleNanoarray_Study01_RTATable.txt

11. 90degpol_RectangleNanoarray_Study02_RTATable.txt

12. 90degpol_RectangleNanoarray_Study03_RTATable.txt

13. 90degpol_RectangleNanoarray_Study04_RTATable.txt

File names containing:

- "0degpol" correspond to 0-rad linearly polarised light excitation.

- "30degpol" correspond to π/6-rad linearly polarised light excitation.

- "45degpol" correspond to π/4-rad linearly polarised light excitation.

- "60degpol" correspond to π/3-rad linearly polarised light excitation.

- "90degpol" correspond to π/2-rad linearly polarised light excitation.

- "Study00" correspond to the 247.5-nm-by-247.5-nm square-shaped nanostructure array (1:1 aspect ratio).

- "Study01" correspond to the 221.4-nm-by-276.7-nm rectangle-shaped nanostructure array (1:1.25 aspect ratio).

- "Study02" correspond to the 202.1-nm-by-303.1-nm rectangle-shaped nanostructure array (1:1.5 aspect ratio).

- "Study03" correspond to the 187.1-nm-by-327.4-nm rectangle-shaped nanostructure array (1:1.75 aspect ratio).

- "Study04" correspond to the 175-nm-by-350-nm rectangle-shaped nanostructure array (1:2 aspect ratio).

## Section S8.3: Rectangle-bound t-shaped nanostructure arrays

Seventeen .txt files are provided containing reflectance, transmittance, and absorptance data of rectangle-bound nanostructure arrays excited by varying angles of linearly polarised light:

1. 0degpol_RectangleBoundTShapedNanoarray_Study00_RTATable.txt

2. 0degpol_RectangleBoundTShapedNanoarray_Study01_RTATable.txt

3. 0degpol_RectangleBoundTShapedNanoarray_Study02_RTATable.txt

4. 0degpol_RectangleBoundTShapedNanoarray_Study03_RTATable.txt

5. 30degpol_RectangleBoundTShapedNanoarray_Study00_RTATable.txt

6. 30degpol_RectangleBoundTShapedNanoarray_Study01_RTATable.txt

7. 30degpol_RectangleBoundTShapedNanoarray_Study03_RTATable.txt

8. 45degpol_RectangleBoundTShapedNanoarray_Study00_RTATable.txt

9. 45degpol_RectangleBoundTShapedNanoarray_Study01_RTATable.txt

10. 45degpol_RectangleBoundTShapedNanoarray_Study03_RTATable.txt

11. 60degpol_RectangleBoundTShapedNanoarray_Study00_RTATable.txt

12. 60degpol_RectangleBoundTShapedNanoarray_Study01_RTATable.txt

13. 60degpol_RectangleBoundTShapedNanoarray_Study03_RTATable.txt

14. 90degpol_RectangleBoundTShapedNanoarray_Study00_RTATable.txt

15. 90degpol_RectangleBoundTShapedNanoarray_Study01_RTATable.txt

16. 90degpol_RectangleBoundTShapedNanoarray_Study02_RTATable.txt

17. 90degpol_RectangleBoundTShapedNanoarray_Study03_RTATable.txt

File names containing:

- "0degpol" correspond to 0-rad linearly polarised light excitation.

- "30degpol" correspond to $\pi/6$-rad linearly polarised light excitation.

- "45degpol" correspond to $\pi/4$-rad linearly polarised light excitation.

- "60degpol" correspond to $\pi/3$-rad linearly polarised light excitation.

- "90degpol" correspond to $\pi/2$-rad linearly polarised light excitation.

- "Study00" correspond to the 327.4-nm-by-247.5-nm rectangle-shaped nanostructure array. This nanostructure array acts as the bound for the remaining t-shaped structures in this folder.

- "Study01" correspond to the rectangle-bound 247.5 nm & 187.1 nm arms thicknesses t-shaped nanostructure array.

- "Study02" correspond to the rectangle-bound 247.5 nm & 143.0 nm arms thicknesses t-shaped nanostructure array.

- "Study03" correspond to the rectangle-bound 247.5 nm & 90.7 nm arms thicknesses t-shaped nanostructure array.

**Section S8.4: Square-bound t-shaped nanostructure arrays**

Nineteen .txt files are provided containing reflectance, transmittance, and absorptance data of square-bound nanostructure arrays excited by varying angles of linearly polarised light:

1. 0degpol_SquareBoundTShapedNanoarray_Study00_RTATable.txt

2. 0degpol_SquareBoundTShapedNanoarray_Study01_RTATable.txt

3. 0degpol_SquareBoundTShapedNanoarray_Study02_RTATable.txt

4. 0degpol_SquareBoundTShapedNanoarray_Study03_RTATable.txt

5. 0degpol_SquareBoundTShapedNanoarray_Study04_RTATable.txt

6. 30degpol_SquareBoundTShapedNanoarray_Study00_RTATable.txt

7. 30degpol_SquareBoundTShapedNanoarray_Study01_RTATable.txt

8. 30degpol_SquareBoundTShapedNanoarray_Study02_RTATable.txt

9. 45degpol_SquareBoundTShapedNanoarray_Study00_RTATable.txt

10. 45degpol_SquareBoundTShapedNanoarray_Study01_RTATable.txt

11. 45degpol_SquareBoundTShapedNanoarray_Study02_RTATable.txt
12. 60degpol_SquareBoundTShapedNanoarray_Study00_RTATable.txt
13. 60degpol_SquareBoundTShapedNanoarray_Study01_RTATable.txt
14. 60degpol_SquareBoundTShapedNanoarray_Study02_RTATable.txt
15. 90degpol_SquareBoundTShapedNanoarray_Study00_RTATable.txt
16. 90degpol_SquareBoundTShapedNanoarray_Study01_RTATable.txt
17. 90degpol_SquareBoundTShapedNanoarray_Study02_RTATable.txt
18. 90degpol_SquareBoundTShapedNanoarray_Study03_RTATable.txt
19. 90degpol_SquareBoundTShapedNanoarray_Study04_RTATable.txt

File names containing:

- "0degpol" correspond to 0-rad linearly polarised light excitation.
- "30degpol" correspond to $\pi/6$-rad linearly polarised light excitation.
- "45degpol" correspond to $\pi/4$-rad linearly polarised light excitation.
- "60degpol" correspond to $\pi/3$-rad linearly polarised light excitation.
- "90degpol" correspond to $\pi/2$-rad linearly polarised light excitation.
- "Study00" correspond to the 327.4-nm-by-327.4-nm square-shaped nanostructure array. This nanostructure array acts as the bound for the remaining t-shaped structures in this folder.
- "Study01" correspond to the square-bound 234.4 nm & 234.4 nm arms thicknesses t-shaped nanostructure array.
- "Study02" correspond to the square-bound 234.4 nm & 162.6 nm arms thicknesses t-shaped nanostructure array.
- "Study03" correspond to the square-bound 234.4 nm & 90.7 nm arms thicknesses t-shaped nanostructure array.
- "Study04" correspond to the square-bound 162.6 nm & 162.6 nm arms thicknesses t-shaped nanostructure array.